%% file: main.tex
\documentclass[10pt,journal,compsoc]{IEEEtran}
\ifCLASSOPTIONcompsoc
  \usepackage[nocompress]{cite}
\else
  \usepackage{cite}
\fi
\ifCLASSINFOpdf
\else
\fi
\usepackage{graphicx}
\usepackage{amsmath}
\usepackage[dvipsnames]{xcolor}
\usepackage{soul}
\usepackage[hidelinks]{hyperref}
\usepackage{multirow}
\usepackage{adjustbox}
\usepackage{tabu}                      
\usepackage{booktabs}                  
\usepackage{makecell}

\begin{document}

%


\title{Real-and-Present: Investigating the Use of Life-Size 2D Video Avatars in HMD-Based AR Teleconferencing}
%
%
%
%

\author{Xuanyu Wang, Weizhan Zhang, Christian Sandor, and Hongbo Fu
\IEEEcompsocitemizethanks
{
\IEEEcompsocthanksitem Corresponding authors: Hongbo Fu and Weizhan Zhang.
\IEEEcompsocthanksitem Xuanyu Wang is with the BDKE Lab, School of Computer Science and Technology, Xi'an Jiaotong University and the School of Creative Media, City University of Hong Kong. E-mail: xwang2247-c@my.cityu.edu.hk.
\IEEEcompsocthanksitem Weizhan Zhang is with the BDKE Lab, School of Computer Science and Technology, Xi'an Jiaotong University. E-mail: zhangwzh@xjtu.edu.cn.
\IEEEcompsocthanksitem Christian Sandor is with Université Paris-Saclay / CNRS, Laboratoire Interdisciplinaire des Sciences du Numérique (LISN). E-mail: chris.sandor@gmail.com.
\IEEEcompsocthanksitem Hongbo Fu is with the School of Creative Media, City University of Hong Kong. E-mail: hongbofu@cityu.edu.hk.

}
\thanks{Manuscript received xx xx, xxxx; revised xx xx, xxxx.}
}

%
%

\markboth{}%
{}
%



\IEEEtitleabstractindextext{%
\begin{abstract}
Augmented Reality (AR) teleconferencing allows separately located users to interact with each other in 3D through agents in their own physical environments. Existing methods leveraging volumetric capturing and reconstruction can provide a high-fidelity experience but are often too complex and expensive for everyday usage. Other solutions target mobile and effortless-to-setup teleconferencing on AR Head Mounted Displays (HMD). {They directly transplant} the conventional video conferencing onto an AR-HMD platform or {use} avatars to represent remote participants. However, they can only support either {a high fidelity or a high level of co-presence}. Moreover, the limited Field of View (FoV) of HMDs could further influence users' immersive experience. To achieve a balance between fidelity and co-presence, we explore using life-size 2D video-based avatars (video avatars for short) in AR teleconferencing. Specifically, with the potential effect of FoV on users' perception of proximity, we first conduct a pilot study to explore the local-user-centered optimal placement of video avatars in small-group AR conversations. With the placement results, we then implement a proof-of-concept prototype of video-avatar-based teleconferencing. We conduct user evaluations with the prototype to verify its effectiveness in balancing fidelity and co-presence. Following the indication in the pilot study, we further {quantitatively} explore the effect of FoV size on the video avatar's optimal placement through a user study involving more FoV conditions in a VR-simulated environment. {We regress placement models to} serve as references for {computationally determining video avatar placements in} such teleconferencing applications on various existing AR HMDs and future ones with bigger FoVs.
\end{abstract}

}

\maketitle

\IEEEdisplaynontitleabstractindextext

%
\IEEEpeerreviewmaketitle

\input{introduction.tex}

\input{relatedwork.tex}
\input{pilotstudy.tex}
\input{evaluation.tex}
\input{vrstudy.tex}
\input{conclusion.tex}
\ifCLASSOPTIONcaptionsoff
  \newpage
\fi



\bibliographystyle{IEEEtran}
\bibliography{ref_file}
%



%

\begin{IEEEbiography}[{\includegraphics[width=1in,height=1.25in,clip,keepaspectratio]{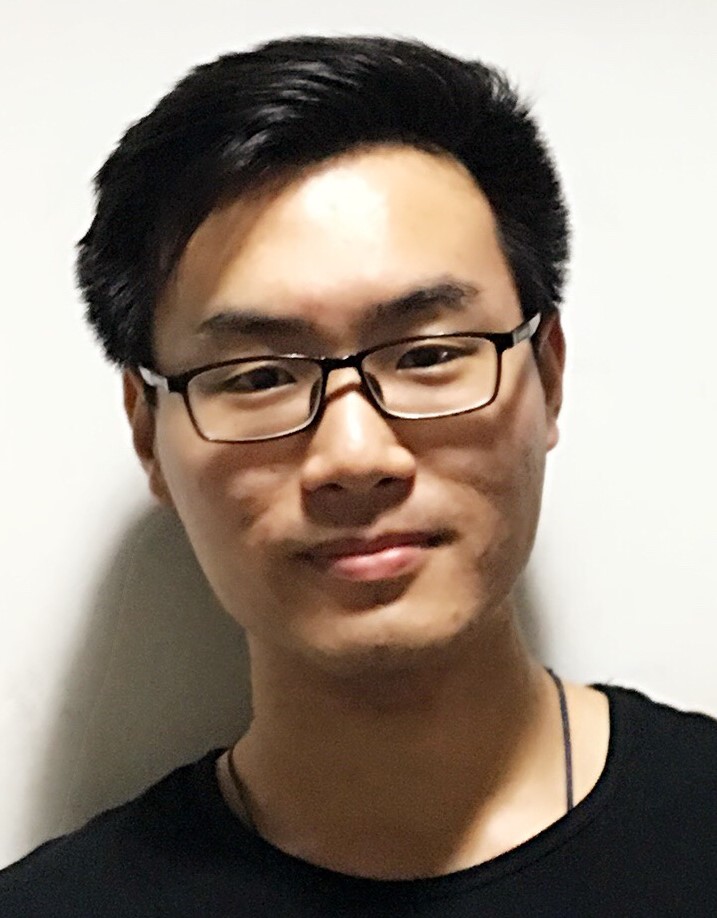}}]{Xuanyu Wang}
received a bachelor's degree in computer science and technology from Xi'an Jiaotong University in 2018. He is currently working towards a joint PhD degree at the School of Computer Science and Technology, Xi'an Jiaotong University and the School of Creative Media, City University of Hong Kong. His research interests lie in the intersection between Human-Computer Interaction and Computer Graphics.
\end{IEEEbiography}
\begin{IEEEbiography}[{\includegraphics[width=1in,height=1.25in,clip,keepaspectratio]{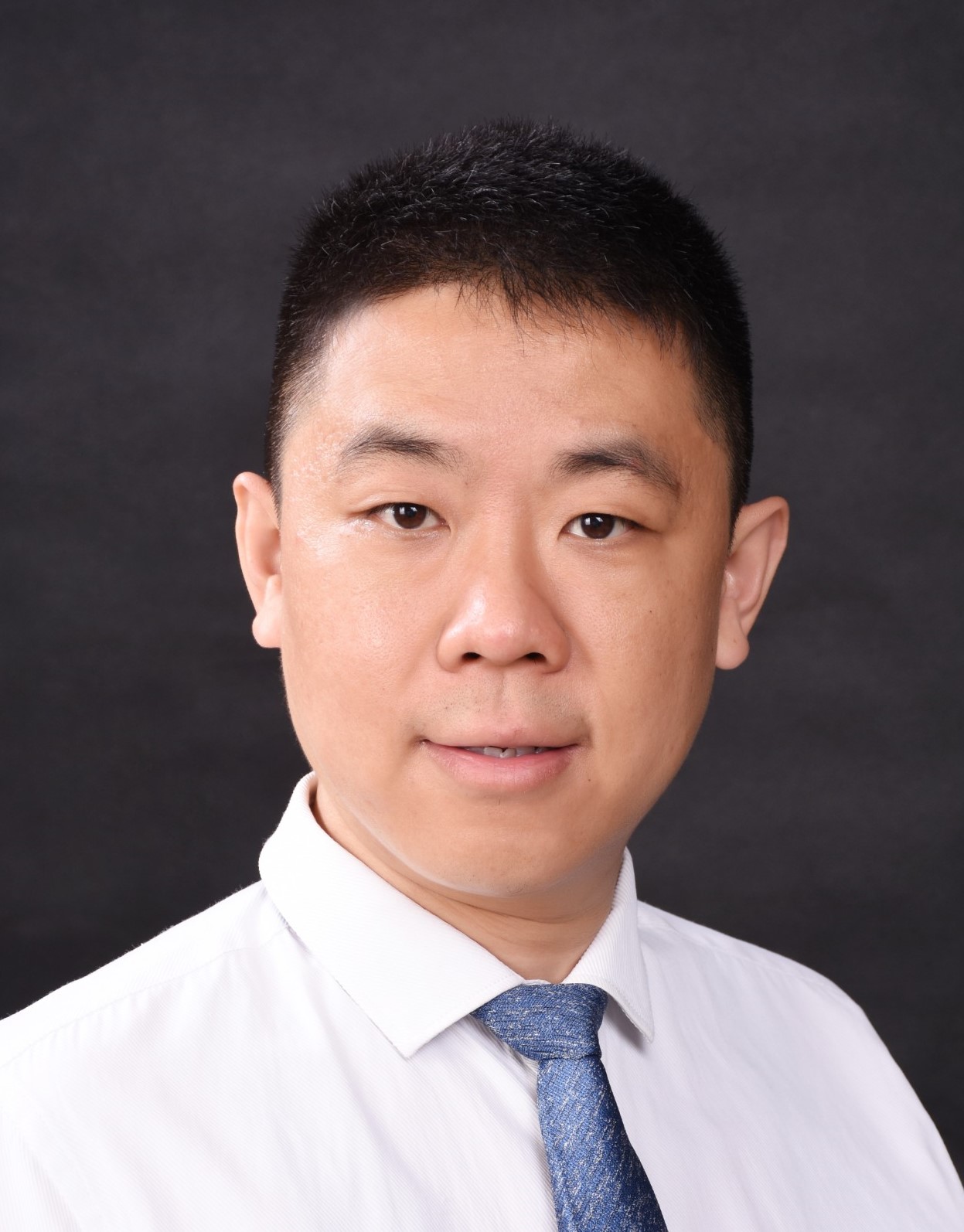}}]{Weizhan Zhang} received the BS degree in electronics engineering from Zhejiang University, China, in 1999, and the PhD degree in computer science from Xi’an Jiaotong University, China, in 2010. He was also a visiting scholar in the Department of Computer Science and Engineering at Pennsylvania State University, in 2015. Now he is a professor at the School of Computer Science and Technology, Xi’an Jiaotong University. His research interests include multimedia networking, edge intelligence, and mobile computing.
\end{IEEEbiography}
\begin{IEEEbiography}[{\includegraphics[width=1in,height=1.25in,clip,keepaspectratio]{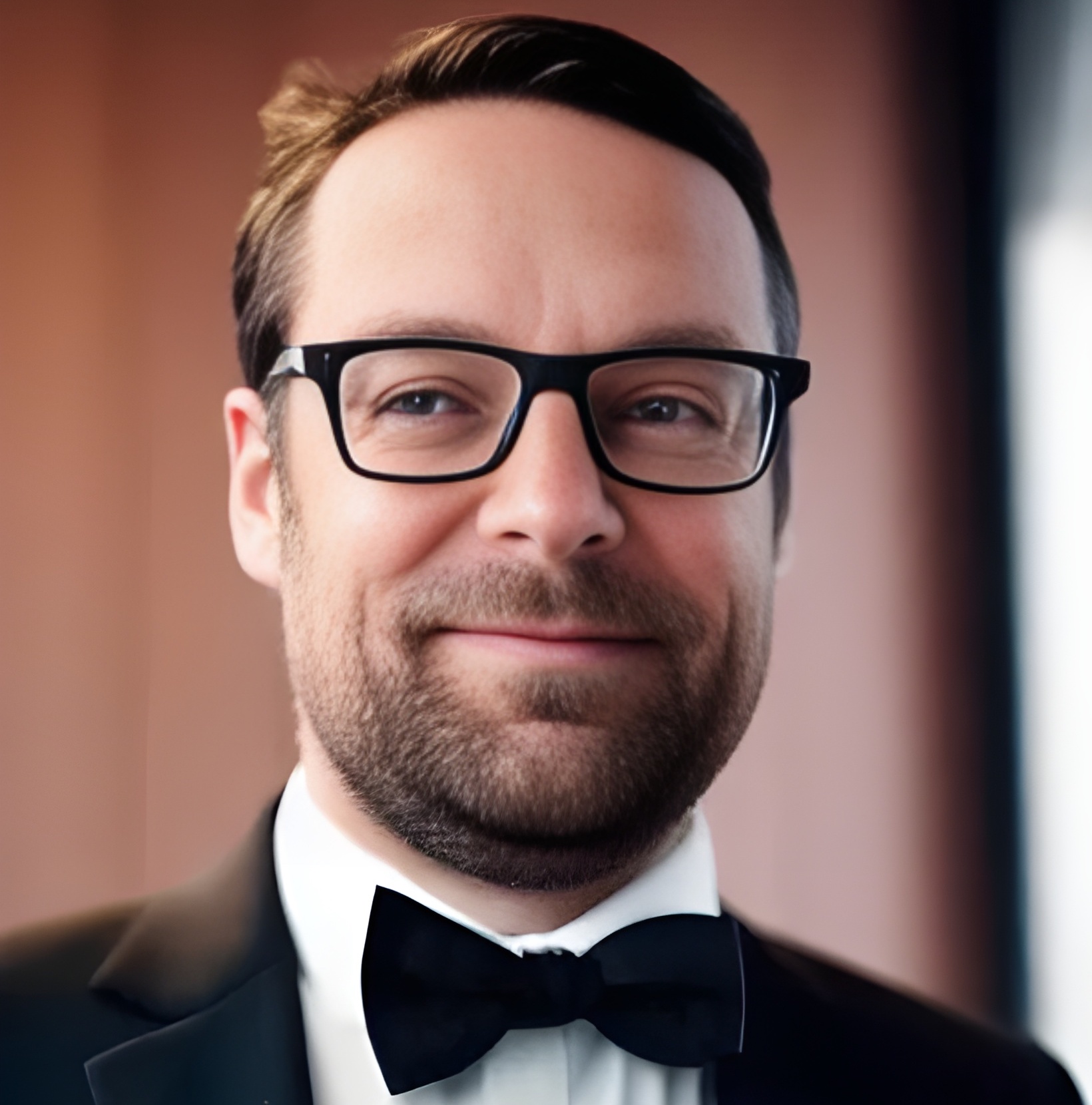}}]{Christian Sandor} received the doctorate degree in computer science from the Munich University of Technology, Germany, in 2005 under the supervision of Prof. Gudrun Klinker and Prof. Steven Feiner. He is a Professor at Université Paris-Saclay and the leader of the VENISE team at CNRS (Centre National de la Recherche Scientifique). He is a member of the IEEE. He serves as an editorial board member for IEEE Transactions on Visualization and Computer Graphics and as a steering committee member for ACM Symposium on Spatial User Interaction and IEEE ISMAR (deputy chair). He has been program chair for numerous conferences, including IEEE ISMAR, SIGGRAPH Asia XR, and SIGGRAPH Asia Symposium On Mobile Graphics And Interactive Applications.
\end{IEEEbiography}
\begin{IEEEbiography}[{\includegraphics[width=1in,height=1.25in,clip,keepaspectratio]{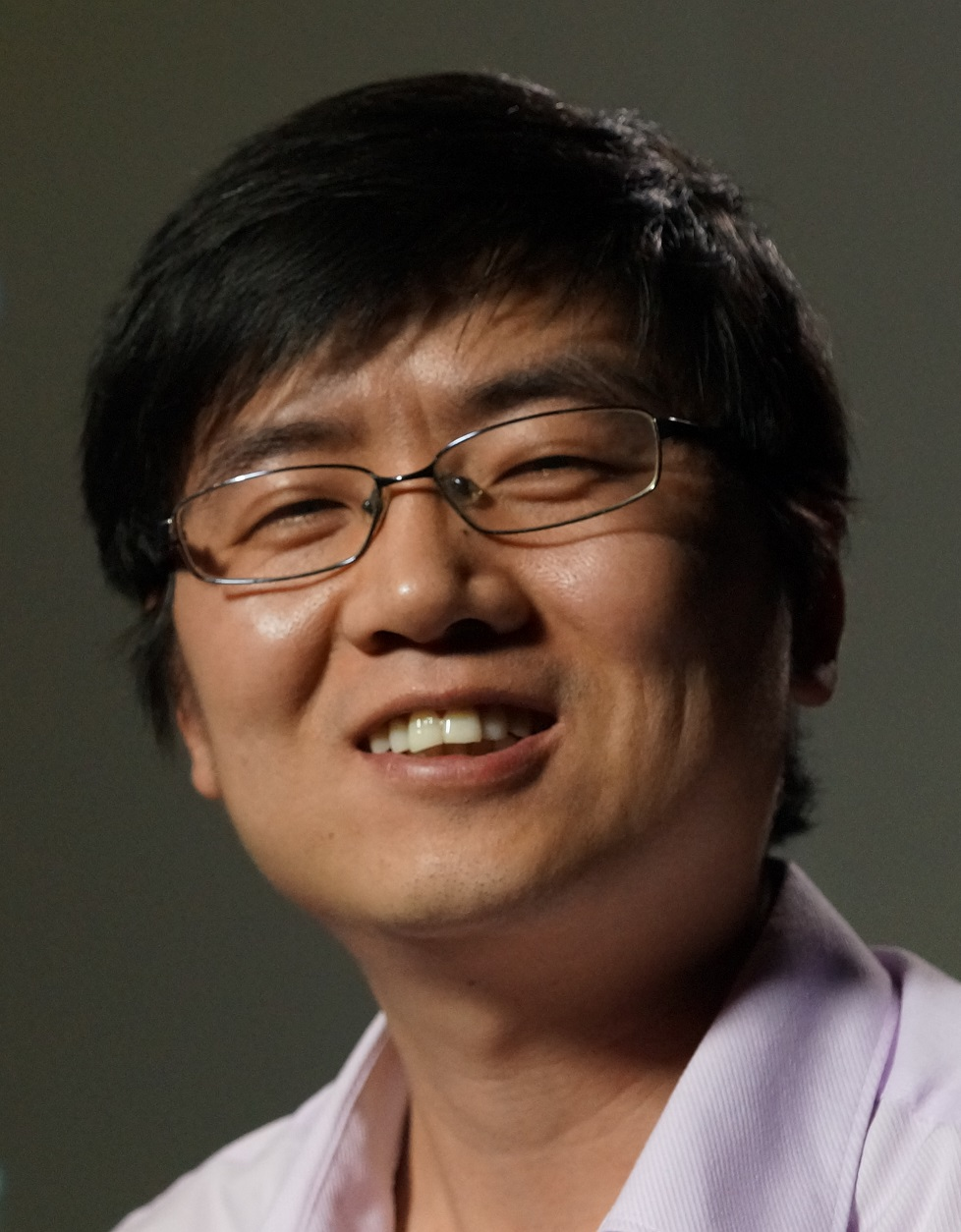}}]{Hongbo Fu}
received a BS degree in information sciences in 2002 from Peking University, China and a PhD degree in computer science from the Hong Kong University of Science and Technology in 2007. He is a Professor at the School of Creative Media, City University of Hong Kong. His primary research interests fall in the fields of computer graphics and human computer interaction. He has served as an Associate Editor of The Visual Computer, Computers \& Graphics, and Computer Graphics Forum.
\end{IEEEbiography}







\end{document}


%


\title{Real-and-Present: Investigating the Use of Life-Size 2D Video Avatars in HMD-Based AR Teleconferencing}
%
%
%
%

\author{Xuanyu Wang, Weizhan Zhang, Christian Sandor, and Hongbo Fu\\Supplemental File
\thanks{Manuscript received xx xx, xxxx; revised xx xx, xxxx.}
}

%
%

\markboth{IEEE TRANSACTIONS ON VISUALIZATION AND COMPUTER GRAPHICS}%
{IEEE TRANSACTIONS ON VISUALIZATION AND COMPUTER GRAPHICS}
\maketitle

\IEEEdisplaynontitleabstractindextext

%
\IEEEpeerreviewmaketitle

\IEEEraisesectionheading{\section{Overview about the Supplemental File}}
Here in the Supplemental File, we report the optimal placement for the \textbf{30-} and \textbf{40-FoV} conditions in the pilot study in Sec. \ref{PilotStudyOptimalPlacement}, report the correlation test results in the pilot study in Sec. \ref{PilotStudyCorrelation}, present the AR {and} VR comparison results in the further VR study in Sec. \ref{ARVRComparison}, and present the Aspect Ratio Comparison in the further VR study in Sec. \ref{AspectRatioComparison}.

\section{Pilot Study}
\subsection{The optimal placement for the \textbf{30-} and \textbf{40-FoV} conditions}\label{PilotStudyOptimalPlacement}
For the \textbf{30-FoV} condition, the optimal $(Radian, Radius)$ is $(/, 1.24)$ for the 1-RU scenario, $(33.75, 1.26)$ for the 2-RU scenario, $(59.64, 1.31)$ for the 3-RU scenario, and $(72.97, 1.61)$ for the 4-RU scenario. For the \textbf{40-FoV} condition, they are $(/, 1.17)$ for the 1-RU scenario, $(39.06, 1.20)$ for the 2-RU scenario, $(68.14, 1.21)$ for the 3-RU scenario, and $(75.20, 1.55)$ for the 4-RU scenario.

\subsection{Correlation Test Results in the Pilot Study}\label{PilotStudyCorrelation}
We report the correlation test results in the pilot study in Tab. \ref{tab:CorrelationPilotStudy}.

\begin{table}[h]
  \caption{{Correlation Test results between the $Radian$ and $Radius$ values and the $FoV$ in the pilot study. ($*$ $p < .05$, $**$ $p < .01$, $***$ $p< .001$)}}
  \label{tab:CorrelationPilotStudy}
  \scriptsize
  \centering
  \begin{tabu}{r*{7}{c}*{2}{r}}
  	\toprule
        &           & 1-RU   & 2-RU   & 3-RU  & 4-RU  \\ 
        \midrule
        \multirow{2}{*}{Radian} & Statistic & $/$  & \begin{tabular}[c]{@{}c@{}}Pearson\\ .24\end{tabular}   & \begin{tabular}[c]{@{}c@{}}Pearson\\ .17\end{tabular}   & \begin{tabular}[c]{@{}c@{}}Spearman\\ .11\end{tabular}  \\
        & P-Value   & $/$  & $.087$    & $.230$     & $.426$   \\
        \multirow{2}{*}{Radius} & Statistic & \begin{tabular}[c]{@{}c@{}}Pearson\\ -.21\end{tabular} & \begin{tabular}[c]{@{}c@{}}Spearman\\ -.21\end{tabular} & \begin{tabular}[c]{@{}c@{}}Spearman\\ -.10\end{tabular} & \begin{tabular}[c]{@{}c@{}}Spearman\\ -.40\end{tabular} \\
        & P-Value   & $.130$   & $.132$     & $.480$       & $.778$   \\
  	\bottomrule
  \end{tabu}
\end{table}

\section{VR Study}\label{VRStudy}

\subsection{AR {and} VR Comparison}\label{ARVRComparison}
\begin{figure*}[htbp]
  \centering
  \includegraphics[width=\textwidth]{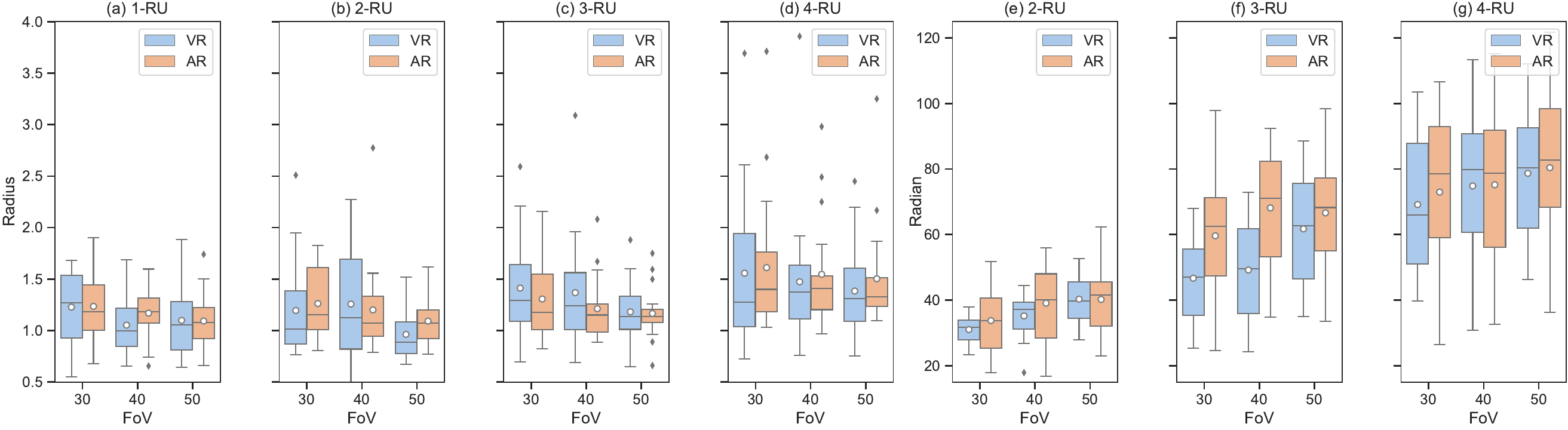}
  \caption{%
  	The statistical results for the data comparison in AR ad VR environments in \textbf{30-}, \textbf{40-}, and \textbf{50-FoV} conditions. (a - d) are the Radius data in 1-, 2-, 3-, and 4-RU scenarios, and they share the y-axis to be the Radius values (as labeled on the left of (a)). 
   (e - g) are the Radian data in 2-, 3-, and 4-RU scenarios, and they share the y-axis to be the Radian values (as labeled on the left of (e)). 
  }
  \label{fig:ARVRComparison}
\end{figure*}

We show the comparison of the data from the further VR study and the pilot study in \textbf{30-}, \textbf{40-}, and \textbf{50-FoV} conditions with the 3:2 aspect ratio in Fig. \ref{fig:ARVRComparison} and report the significance test {and equivalence test} results in Tab. \ref{tab:ARVRComparison}.

\begin{table*}[thpb]
  \caption{{The statistical results for the significance test and equivalence test between data from AR and VR environments. ($*$ $p < .05$, $**$ $p < .01$, $***$ $p < .001$)}}
  \label{tab:ARVRComparison}
\begin{adjustbox}{width=\textwidth}
  \centering
\begin{tabular}{|lll|lll|lll|lll|lll|}
\hline
\multicolumn{2}{|l|}{\multirow{2}{*}{}}                                                  & Scenario & \multicolumn{3}{c|}{1-RU}                                           & \multicolumn{3}{c|}{2-RU}                                           & \multicolumn{3}{c|}{3-RU}                                           & \multicolumn{3}{c|}{4-RU}                                          \\ \cline{3-15} 
\multicolumn{2}{|l|}{}                                                                   & FoV      & \multicolumn{1}{l|}{30}     & \multicolumn{1}{l|}{40}     & 50      & \multicolumn{1}{l|}{30}     & \multicolumn{1}{l|}{40}     & 50      & \multicolumn{1}{l|}{30}     & \multicolumn{1}{l|}{40}     & 50      & \multicolumn{1}{l|}{30}     & \multicolumn{1}{l|}{40}     & 50     \\ \hline
\multicolumn{1}{|l|}{\multirow{6}{*}{Radian}} & \multicolumn{1}{l|}{\multirow{2}{*}{AR}} & M        & \multicolumn{1}{l|}{$/$}    & \multicolumn{1}{l|}{$/$}    & $/$     & \multicolumn{1}{l|}{33.8}  & \multicolumn{1}{l|}{39.1}  & 40.2   & \multicolumn{1}{l|}{59.6}  & \multicolumn{1}{l|}{68.1}  & 66.6   & \multicolumn{1}{l|}{73.0}  & \multicolumn{1}{l|}{75.2}  & 80.4  \\ \cline{3-15} 
\multicolumn{1}{|l|}{}                        & \multicolumn{1}{l|}{}                    & SD       & \multicolumn{1}{l|}{$/$}    & \multicolumn{1}{l|}{$/$}    & $/$     & \multicolumn{1}{l|}{10.4}  & \multicolumn{1}{l|}{11.7}  & 10.4   & \multicolumn{1}{l|}{18.3}  & \multicolumn{1}{l|}{16.8}  & 15.0   & \multicolumn{1}{l|}{23.6}  & \multicolumn{1}{l|}{23.3}  & 23.1  \\ \cline{2-15} 
\multicolumn{1}{|l|}{}                        & \multicolumn{1}{l|}{\multirow{2}{*}{VR}} & M        & \multicolumn{1}{l|}{$/$}    & \multicolumn{1}{l|}{$/$}    & $/$     & \multicolumn{1}{l|}{31.0}  & \multicolumn{1}{l|}{35.2}  & 40.3   & \multicolumn{1}{l|}{46.7}  & \multicolumn{1}{l|}{49.2}  & 61.8   & \multicolumn{1}{l|}{69.2}  & \multicolumn{1}{l|}{74.8}  & 78.7  \\ \cline{3-15} 
\multicolumn{1}{|l|}{}                        & \multicolumn{1}{l|}{}                    & SD       & \multicolumn{1}{l|}{$/$}    & \multicolumn{1}{l|}{$/$}    & $/$     & \multicolumn{1}{l|}{4.4}   & \multicolumn{1}{l|}{6.3}   & 7.2    & \multicolumn{1}{l|}{13.3}  & \multicolumn{1}{l|}{14.8}  & 16.7   & \multicolumn{1}{l|}{20.3}  & \multicolumn{1}{l|}{20.40}  & 18.8  \\ \cline{2-15} 
\multicolumn{1}{|l|}{}                        & \multicolumn{2}{l|}{Statistic}                      & \multicolumn{1}{l|}{$/$}    & \multicolumn{1}{l|}{$/$}    & $/$     & \multicolumn{1}{l|}{z = 67.0} & \multicolumn{1}{l|}{z = 61.0} & t = -0.03 & \multicolumn{1}{l|}{t = 2.36} & \multicolumn{1}{l|}{t = 3.50} & t = 0.90  & \multicolumn{1}{l|}{t = 0.50} & \multicolumn{1}{l|}{t = 0.05} & t = 0.24 \\ \cline{2-15} 
\multicolumn{1}{|l|}{}                        & \multicolumn{2}{l|}{P-Value}                        & \multicolumn{1}{l|}{$/$}    & \multicolumn{1}{l|}{$/$}    & $/$     & \multicolumn{1}{l|}{.442}   & \multicolumn{1}{l|}{.304}   & .979    & \multicolumn{1}{l|}{\color{orange}$*$}      & \multicolumn{1}{l|}{\color{orange}$**$}     & .375    & \multicolumn{1}{l|}{.618}   & \multicolumn{1}{l|}{.962}   & .813   \\ \cline{2-15}
\multicolumn{1}{|l|}{}                        & \multicolumn{2}{l|}{{TOST P-Value}}                        & \multicolumn{1}{l|}{$/$}    & \multicolumn{1}{l|}{$/$}    & $/$     & \multicolumn{1}{l|}{$***$}   & \multicolumn{1}{l|}{$***$}   & $***$    & \multicolumn{1}{l|}{.106}      & \multicolumn{1}{l|}{.418}     & $**$    & \multicolumn{1}{l|}{$*$}   & \multicolumn{1}{l|}{$**$}   & $*$   \\ \hline
\multicolumn{1}{|l|}{\multirow{6}{*}{Radius}} & \multicolumn{1}{l|}{\multirow{2}{*}{AR}} & M        & \multicolumn{1}{l|}{1.2}   & \multicolumn{1}{l|}{1.2}   & 1.1    & \multicolumn{1}{l|}{1.3}   & \multicolumn{1}{l|}{1.2}   & 1.1    & \multicolumn{1}{l|}{1.3}   & \multicolumn{1}{l|}{1.2}   & 1.2    & \multicolumn{1}{l|}{1.6}   & \multicolumn{1}{l|}{1.5}   & 1.5   \\ \cline{3-15} 
\multicolumn{1}{|l|}{}                        & \multicolumn{1}{l|}{}                    & SD       & \multicolumn{1}{l|}{0.3}   & \multicolumn{1}{l|}{0.2}   & 0.3    & \multicolumn{1}{l|}{0.3}   & \multicolumn{1}{l|}{0.4}   & 0.2    & \multicolumn{1}{l|}{0.4}   & \multicolumn{1}{l|}{0.3}   & 0.2    & \multicolumn{1}{l|}{0.7}   & \multicolumn{1}{l|}{0.5}   & 0.5   \\ \cline{2-15} 
\multicolumn{1}{|l|}{}                        & \multicolumn{1}{l|}{\multirow{2}{*}{VR}} & M        & \multicolumn{1}{l|}{1.2}   & \multicolumn{1}{l|}{1.1}   & 1.1    & \multicolumn{1}{l|}{1.2}   & \multicolumn{1}{l|}{1.3}   & 1.0    & \multicolumn{1}{l|}{1.4}   & \multicolumn{1}{l|}{1.4}   & 1.2    & \multicolumn{1}{l|}{1.6}   & \multicolumn{1}{l|}{1.5}   & 1.4   \\ \cline{3-15} 
\multicolumn{1}{|l|}{}                        & \multicolumn{1}{l|}{}                    & SD       & \multicolumn{1}{l|}{0.3}   & \multicolumn{1}{l|}{0.3}   & 0.3    & \multicolumn{1}{l|}{0.4}   & \multicolumn{1}{l|}{0.5}   & 0.3    & \multicolumn{1}{l|}{0.5}   & \multicolumn{1}{l|}{0.5}   & 0.3    & \multicolumn{1}{l|}{0.7}   & \multicolumn{1}{l|}{0.7}   & 0.4   \\ \cline{2-15} 
\multicolumn{1}{|l|}{}                        & \multicolumn{2}{l|}{Statistic}                      & \multicolumn{1}{l|}{t = 0.07} & \multicolumn{1}{l|}{t = 1.26} & t = -0.06 & \multicolumn{1}{l|}{z = 57.0} & \multicolumn{1}{l|}{z = 77.0} & z = 48.0  & \multicolumn{1}{l|}{z = 70.0} & \multicolumn{1}{l|}{z = 66.0} & t = -0.17 & \multicolumn{1}{l|}{z = 58.0} & \multicolumn{1}{l|}{z = 61.0} & z = 51.0 \\ \cline{2-15} 
\multicolumn{1}{|l|}{}                        & \multicolumn{2}{l|}{P-Value}                        & \multicolumn{1}{l|}{.945}   & \multicolumn{1}{l|}{.216}   & .956    & \multicolumn{1}{l|}{.229}   & \multicolumn{1}{l|}{.734}   & .108    & \multicolumn{1}{l|}{.523}   & \multicolumn{1}{l|}{.417}   & .866    & \multicolumn{1}{l|}{.246}   & \multicolumn{1}{l|}{.304}   & .142   \\ \cline{2-15}
\multicolumn{1}{|l|}{}                        & \multicolumn{2}{l|}{{TOST P-Value}}                        & \multicolumn{1}{l|}{$***$}    & \multicolumn{1}{l|}{$***$}    & $***$     & \multicolumn{1}{l|}{$***$}   & \multicolumn{1}{l|}{$**$}   & $***$    & \multicolumn{1}{l|}{$**$}      & \multicolumn{1}{l|}{$*$}     & $***$    & \multicolumn{1}{l|}{$**$}   & \multicolumn{1}{l|}{$*$}   & $**$   \\ \hline
\end{tabular}
\end{adjustbox}
\end{table*}

\subsection{Aspect Ratio Comparison}\label{AspectRatioComparison}
\begin{figure}[htbp]
  \centering
  \includegraphics[width=.48\linewidth]{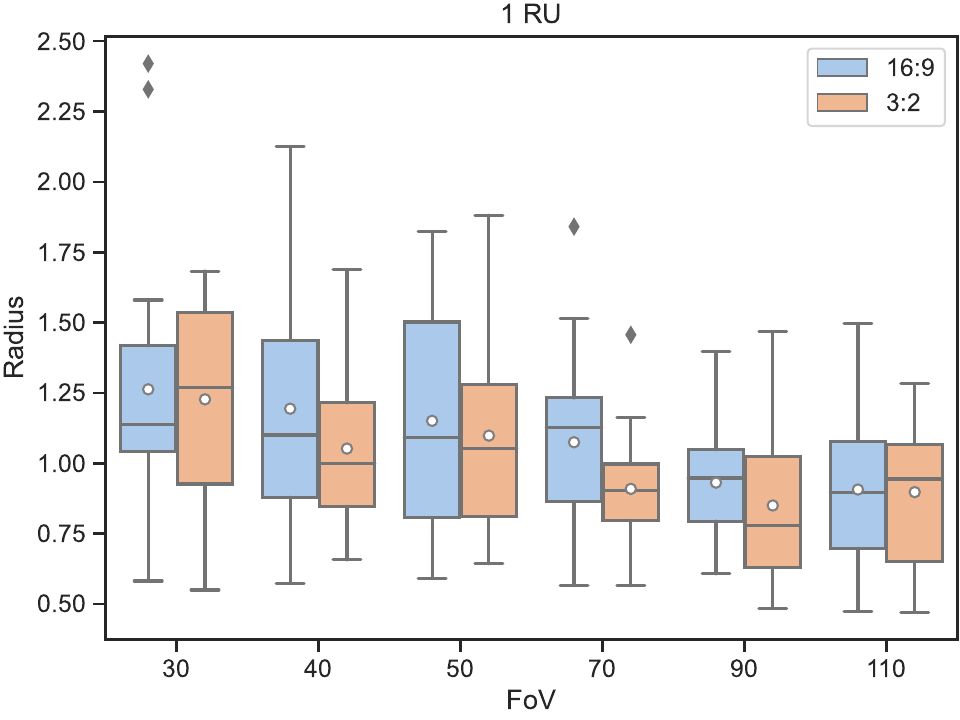}
  \includegraphics[width=.48\linewidth]{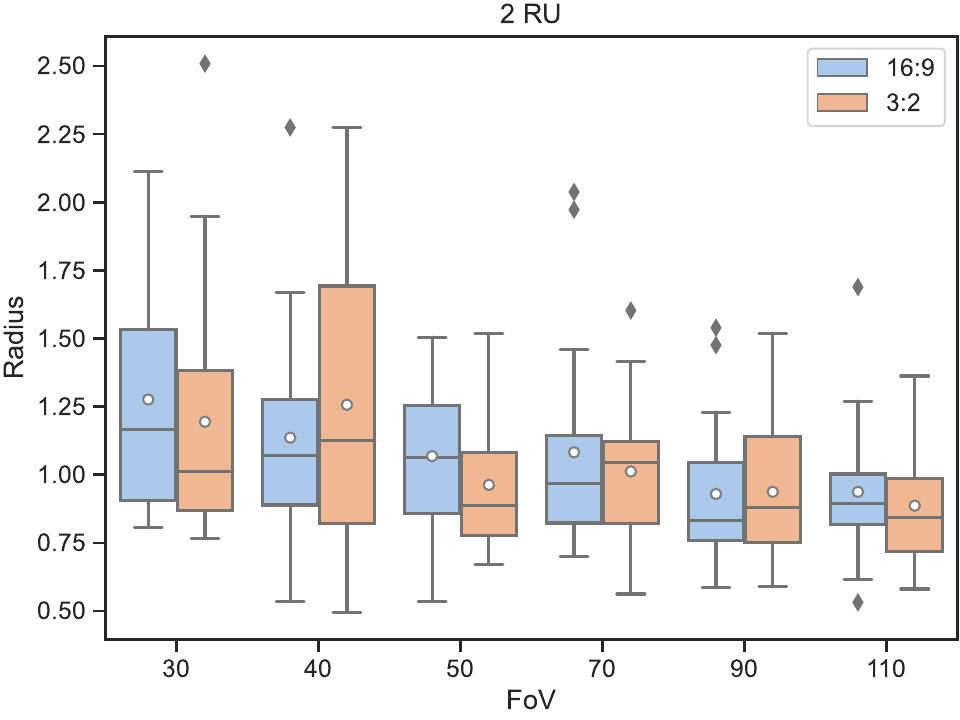}
  \includegraphics[width=.48\linewidth]{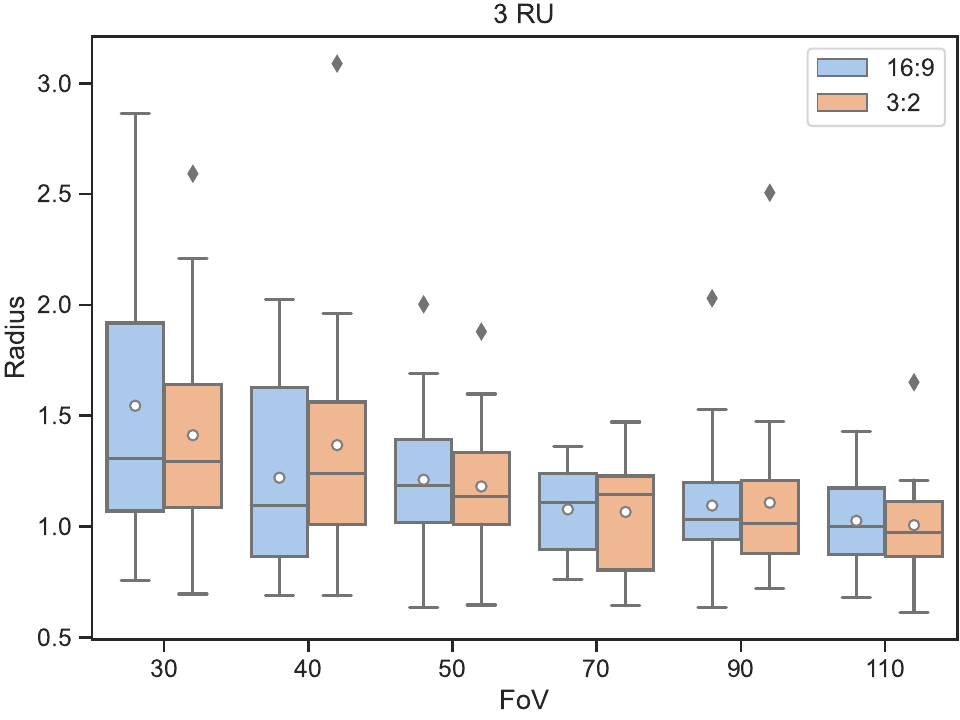}
  \includegraphics[width=.48\linewidth]{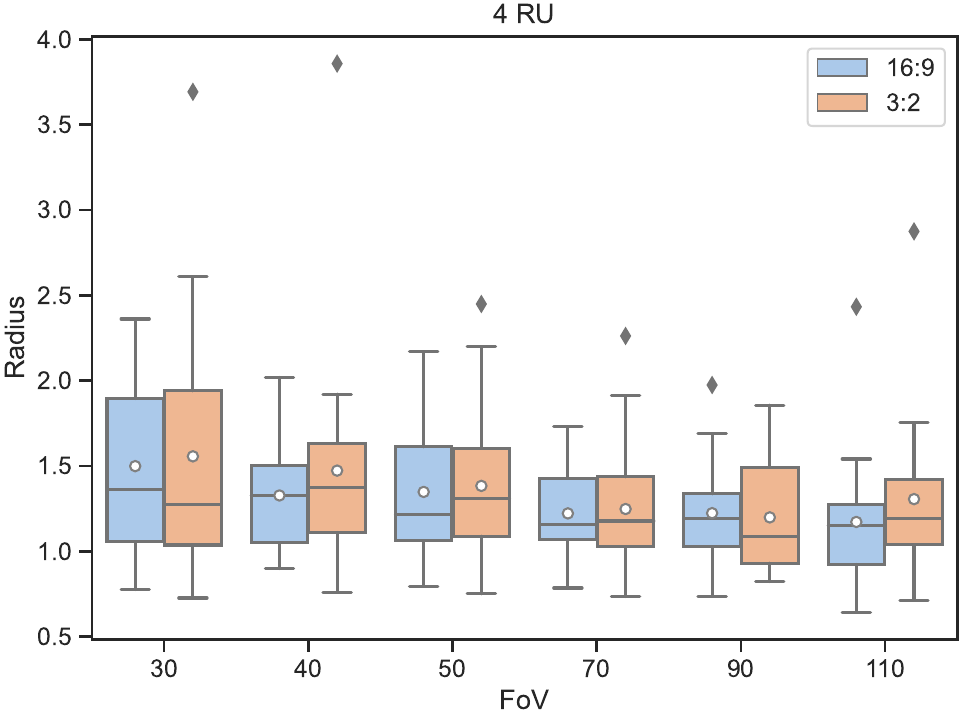}
  \includegraphics[width=.48\linewidth]{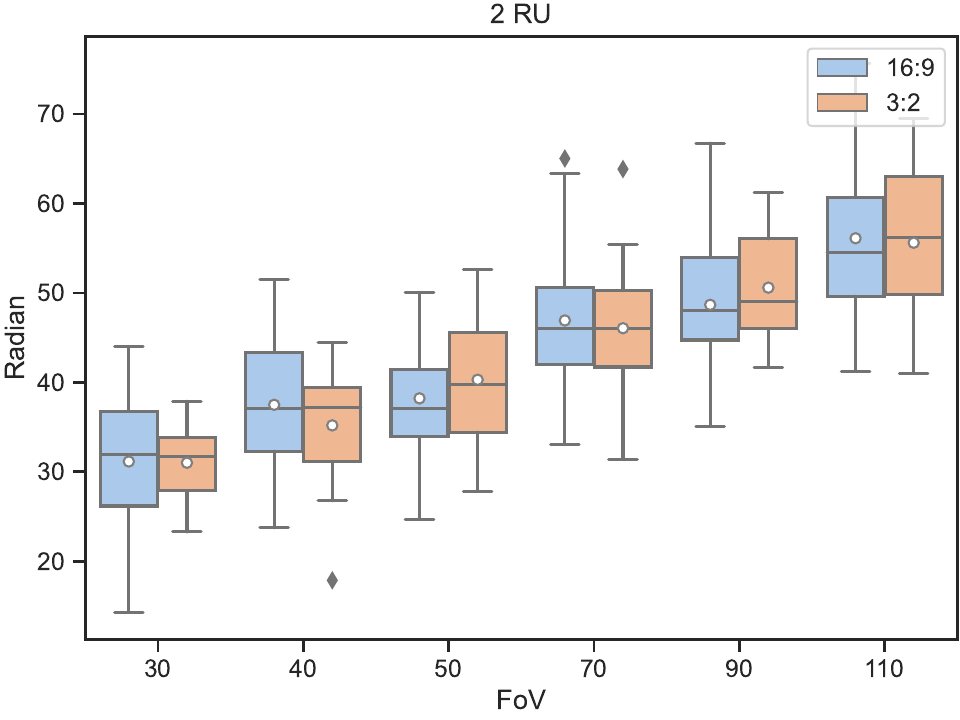}
  \includegraphics[width=.48\linewidth]{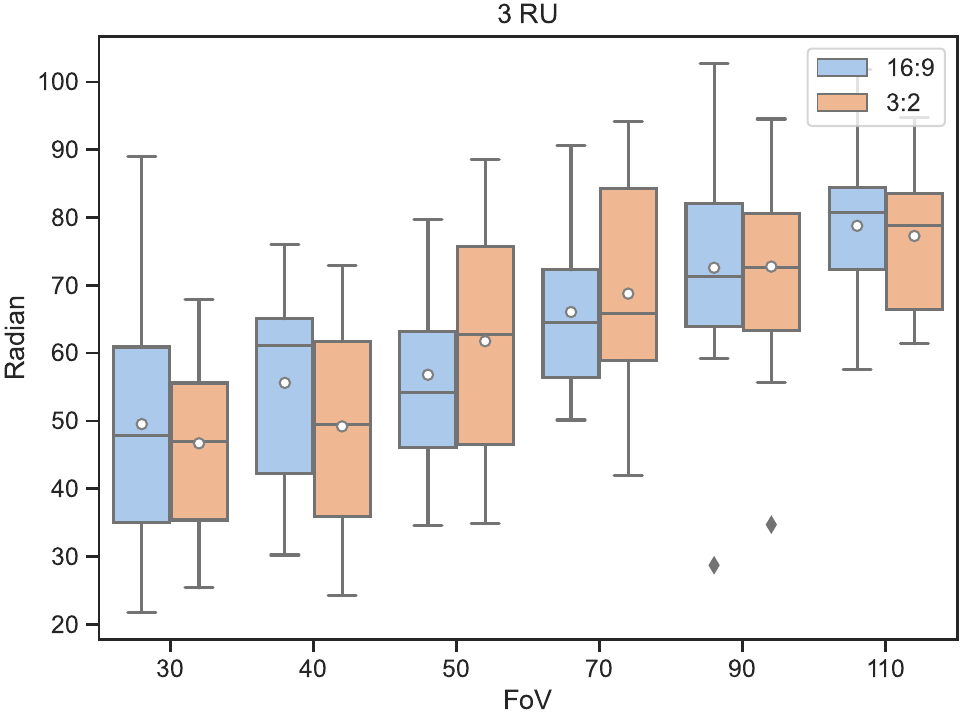}
  \includegraphics[width=.48\linewidth]{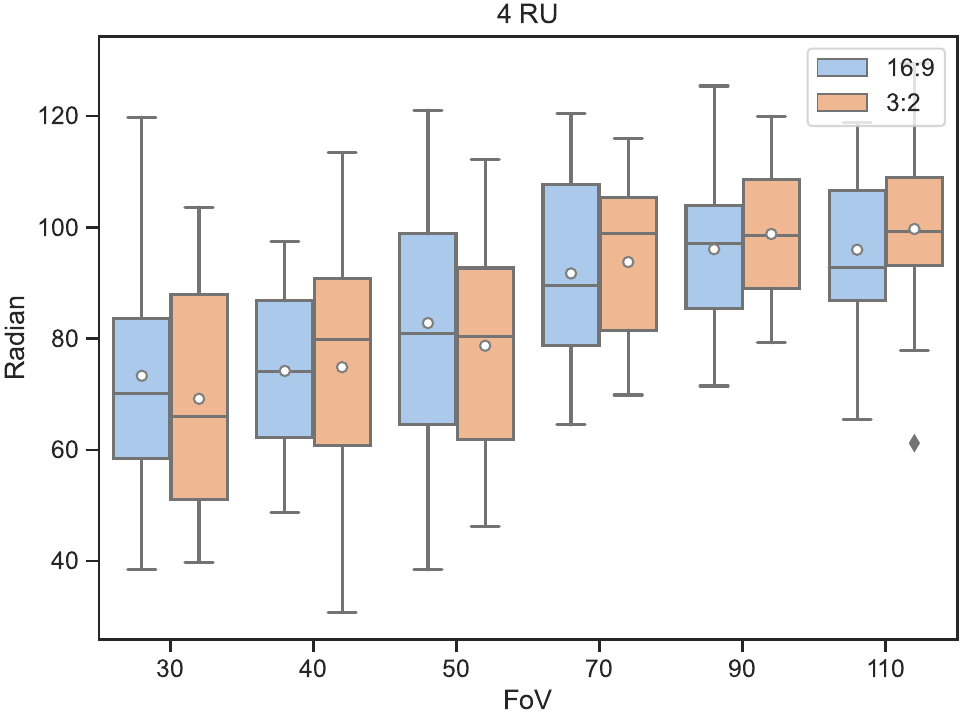}
  \caption{The comparison of the 16:9 and 3:2 aspect ratios in the VR study. {From left to right, in the first two rows} are comparisons of Radius in 1-, 2-, 3-, and 4-RU scenarios, and {in the last two rows} are comparisons of Radian in 2-, 3-, and 4-RU scenarios.}
  \label{fig:AspectRatioComparison}
\end{figure}

We show the comparison between the data from 16:9 and 3:2 aspect ratios in the further VR study in Fig. \ref{fig:AspectRatioComparison} and report the significance test results in Tab. \ref{tab:AspectRatioComparison}.


\begin{table*}[]
  \caption{{The statistical results for the significance test between data from 16:9 and 3:2 aspect ratios. ($*$ $p < .05$, $**$ $p < .01$, $***$ $p < .001$)}}
  \label{tab:AspectRatioComparison}
\begin{adjustbox}{width=\textwidth}
  \centering
\begin{tabular}{|lll|llllll|llllll|}
\hline
\multicolumn{2}{|l|}{\multirow{2}{*}{}}                                                    & Scenario & \multicolumn{6}{c|}{1-RU}                                                                                                                                                                                              & \multicolumn{6}{c|}{2-RU}                                                                                                                                          \\ \cline{3-15} 
\multicolumn{2}{|l|}{}                                                                     & FoV      & \multicolumn{1}{l|}{30}     & \multicolumn{1}{l|}{40}                                 & \multicolumn{1}{l|}{50}      & \multicolumn{1}{l|}{70}                                 & \multicolumn{1}{l|}{90}      & 110    & \multicolumn{1}{l|}{30}      & \multicolumn{1}{l|}{40}      & \multicolumn{1}{l|}{50}      & \multicolumn{1}{l|}{70}      & \multicolumn{1}{l|}{90}      & 110     \\ \hline
\multicolumn{1}{|l|}{\multirow{6}{*}{Radian}} & \multicolumn{1}{l|}{\multirow{2}{*}{16:9}} & M        & \multicolumn{1}{l|}{$/$}    & \multicolumn{1}{l|}{$/$}                                & \multicolumn{1}{l|}{$/$}     & \multicolumn{1}{l|}{$/$}                                & \multicolumn{1}{l|}{$/$}     & $/$    & \multicolumn{1}{l|}{31.2}   & \multicolumn{1}{l|}{37.5}   & \multicolumn{1}{l|}{38.2}   & \multicolumn{1}{l|}{46.9}   & \multicolumn{1}{l|}{48.7}   & 56.1   \\ \cline{3-15} 
\multicolumn{1}{|l|}{}                        & \multicolumn{1}{l|}{}                      & SD       & \multicolumn{1}{l|}{$/$}    & \multicolumn{1}{l|}{$/$}                                & \multicolumn{1}{l|}{$/$}     & \multicolumn{1}{l|}{$/$}                                & \multicolumn{1}{l|}{$/$}     & $/$    & \multicolumn{1}{l|}{7.7}    & \multicolumn{1}{l|}{7.4}    & \multicolumn{1}{l|}{6.2}    & \multicolumn{1}{l|}{8.2}    & \multicolumn{1}{l|}{7.2}    & 8.4    \\ \cline{2-15} 
\multicolumn{1}{|l|}{}                        & \multicolumn{1}{l|}{\multirow{2}{*}{3:2}}  & M        & \multicolumn{1}{l|}{$/$}    & \multicolumn{1}{l|}{$/$}                                & \multicolumn{1}{l|}{$/$}     & \multicolumn{1}{l|}{$/$}                                & \multicolumn{1}{l|}{$/$}     & $/$    & \multicolumn{1}{l|}{31.0}   & \multicolumn{1}{l|}{35.2}   & \multicolumn{1}{l|}{40.3}   & \multicolumn{1}{l|}{46.0}   & \multicolumn{1}{l|}{50.6}   & 55.6   \\ \cline{3-15} 
\multicolumn{1}{|l|}{}                        & \multicolumn{1}{l|}{}                      & SD       & \multicolumn{1}{l|}{$/$}    & \multicolumn{1}{l|}{$/$}                                & \multicolumn{1}{l|}{$/$}     & \multicolumn{1}{l|}{$/$}                                & \multicolumn{1}{l|}{$/$}     & $/$    & \multicolumn{1}{l|}{4.4}    & \multicolumn{1}{l|}{6.3}    & \multicolumn{1}{l|}{7.2}    & \multicolumn{1}{l|}{7.2}    & \multicolumn{1}{l|}{5.8}    & 8.1    \\ \cline{2-15} 
\multicolumn{1}{|l|}{}                        & \multicolumn{2}{l|}{Statistic}                        & \multicolumn{1}{l|}{$/$}    & \multicolumn{1}{l|}{$/$}                                & \multicolumn{1}{l|}{$/$}     & \multicolumn{1}{l|}{$/$}                                & \multicolumn{1}{l|}{$/$}     & $/$    & \multicolumn{1}{l|}{z = 79.0} & \multicolumn{1}{l|}{t = 1.27}  & \multicolumn{1}{l|}{t = -0.93} & \multicolumn{1}{l|}{t = 0.42}  & \multicolumn{1}{l|}{t = -0.96} & t = 0.24  \\ \cline{2-15} 
\multicolumn{1}{|l|}{}                        & \multicolumn{2}{l|}{P-Value}                          & \multicolumn{1}{l|}{$/$}    & \multicolumn{1}{l|}{$/$}                                & \multicolumn{1}{l|}{$/$}     & \multicolumn{1}{l|}{$/$}                                & \multicolumn{1}{l|}{$/$}     & $/$    & \multicolumn{1}{l|}{.799}    & \multicolumn{1}{l|}{.222}    & \multicolumn{1}{l|}{.365}    & \multicolumn{1}{l|}{.678}    & \multicolumn{1}{l|}{.348}    & .815    \\ \hline
\multicolumn{1}{|l|}{\multirow{6}{*}{Radius}} & \multicolumn{1}{l|}{\multirow{2}{*}{16:9}} & M        & \multicolumn{1}{l|}{1.3}   & \multicolumn{1}{l|}{1.2}                               & \multicolumn{1}{l|}{1.2}    & \multicolumn{1}{l|}{1.1}                               & \multicolumn{1}{l|}{0.9}    & 0.9   & \multicolumn{1}{l|}{1.3}    & \multicolumn{1}{l|}{1.1}    & \multicolumn{1}{l|}{1.1}    & \multicolumn{1}{l|}{1.1}    & \multicolumn{1}{l|}{0.9}    & 0.9    \\ \cline{3-15} 
\multicolumn{1}{|l|}{}                        & \multicolumn{1}{l|}{}                      & SD       & \multicolumn{1}{l|}{0.5}   & \multicolumn{1}{l|}{0.4}                               & \multicolumn{1}{l|}{0.4}    & \multicolumn{1}{l|}{0.3}                               & \multicolumn{1}{l|}{0.2}    & 0.3   & \multicolumn{1}{l|}{0.4}    & \multicolumn{1}{l|}{0.4}    & \multicolumn{1}{l|}{0.2}    & \multicolumn{1}{l|}{0.4}    & \multicolumn{1}{l|}{0.3}    & 0.3    \\ \cline{2-15} 
\multicolumn{1}{|l|}{}                        & \multicolumn{1}{l|}{\multirow{2}{*}{3:2}}  & M        & \multicolumn{1}{l|}{1.2}   & \multicolumn{1}{l|}{1.1}                               & \multicolumn{1}{l|}{1.1}    & \multicolumn{1}{l|}{0.9}                               & \multicolumn{1}{l|}{0.9}    & 0.9   & \multicolumn{1}{l|}{1.2}    & \multicolumn{1}{l|}{1.3}    & \multicolumn{1}{l|}{1.0}    & \multicolumn{1}{l|}{1.0}    & \multicolumn{1}{l|}{0.9}    & 0.9    \\ \cline{3-15} 
\multicolumn{1}{|l|}{}                        & \multicolumn{1}{l|}{}                      & SD       & \multicolumn{1}{l|}{0.3}   & \multicolumn{1}{l|}{0.3}                               & \multicolumn{1}{l|}{0.3}    & \multicolumn{1}{l|}{0.2}                               & \multicolumn{1}{l|}{0.3}    & 0.2   & \multicolumn{1}{l|}{0.4}    & \multicolumn{1}{l|}{0.5}    & \multicolumn{1}{l|}{0.3}    & \multicolumn{1}{l|}{0.3}    & \multicolumn{1}{l|}{0.2}    & 0.2    \\ \cline{2-15} 
\multicolumn{1}{|l|}{}                        & \multicolumn{2}{l|}{Statistic}                        & \multicolumn{1}{l|}{z = 82.0} & \multicolumn{1}{l|}{t = 1.56}                             & \multicolumn{1}{l|}{t = 0.74}  & \multicolumn{1}{l|}{t = 2.39}                             & \multicolumn{1}{l|}{t = 1.42}  & t = 0.26 & \multicolumn{1}{l|}{z = 65.0}  & \multicolumn{1}{l|}{t = -1.11} & \multicolumn{1}{l|}{z = 36.0}  & \multicolumn{1}{l|}{z = 77.0}  & \multicolumn{1}{l|}{t = -0.17} & t = 1.16  \\ \cline{2-15} 
\multicolumn{1}{|l|}{}                        & \multicolumn{2}{l|}{P-Value}                          & \multicolumn{1}{l|}{.899}   & \multicolumn{1}{l|}{.137}                               & \multicolumn{1}{l|}{.472}    & \multicolumn{1}{l|}{\color{orange}$*$} & \multicolumn{1}{l|}{.173}    & .800   & \multicolumn{1}{l|}{.393}    & \multicolumn{1}{l|}{.285}    & \multicolumn{1}{l|}{\color{orange}$*$}    & \multicolumn{1}{l|}{.734}    & \multicolumn{1}{l|}{.867}    & .264    \\ \hline
\multicolumn{2}{|l|}{\multirow{2}{*}{}}                                                    & Scenario & \multicolumn{6}{c|}{3-RU}                                                                                                                                                                                              & \multicolumn{6}{c|}{4-RU}                                                                                                                                          \\ \cline{3-15} 
\multicolumn{2}{|l|}{}                                                                     & FoV      & \multicolumn{1}{l|}{30}     & \multicolumn{1}{l|}{40}                                 & \multicolumn{1}{l|}{50}      & \multicolumn{1}{l|}{70}                                 & \multicolumn{1}{l|}{90}      & 110    & \multicolumn{1}{l|}{30}      & \multicolumn{1}{l|}{40}      & \multicolumn{1}{l|}{50}      & \multicolumn{1}{l|}{70}      & \multicolumn{1}{l|}{90}      & 110     \\ \hline
\multicolumn{1}{|l|}{\multirow{6}{*}{Radian}} & \multicolumn{1}{l|}{\multirow{2}{*}{16:9}} & M        & \multicolumn{1}{l|}{49.5}  & \multicolumn{1}{l|}{55.6}                              & \multicolumn{1}{l|}{56.8}   & \multicolumn{1}{l|}{66.1}                              & \multicolumn{1}{l|}{72.6}   & 78.8  & \multicolumn{1}{l|}{73.3}   & \multicolumn{1}{l|}{74.2}   & \multicolumn{1}{l|}{82.8}   & \multicolumn{1}{l|}{91.7}   & \multicolumn{1}{l|}{96.1}   & 96.0   \\ \cline{3-15} 
\multicolumn{1}{|l|}{}                        & \multicolumn{1}{l|}{}                      & SD       & \multicolumn{1}{l|}{17.7}  & \multicolumn{1}{l|}{13.9}                              & \multicolumn{1}{l|}{14.0}   & \multicolumn{1}{l|}{12.2}                              & \multicolumn{1}{l|}{15.6}   & 10.8  & \multicolumn{1}{l|}{22.8}   & \multicolumn{1}{l|}{14.6}   & \multicolumn{1}{l|}{21.9}   & \multicolumn{1}{l|}{17.1}   & \multicolumn{1}{l|}{15.9}   & 14.1   \\ \cline{2-15} 
\multicolumn{1}{|l|}{}                        & \multicolumn{1}{l|}{\multirow{2}{*}{3:2}}  & M        & \multicolumn{1}{l|}{46.7}  & \multicolumn{1}{l|}{49.2}                              & \multicolumn{1}{l|}{61.7}   & \multicolumn{1}{l|}{68.8}                              & \multicolumn{1}{l|}{72.8}   & 77.3  & \multicolumn{1}{l|}{69.2}   & \multicolumn{1}{l|}{74.8}   & \multicolumn{1}{l|}{78.7}   & \multicolumn{1}{l|}{93.8}   & \multicolumn{1}{l|}{98.8}   & 99.7   \\ \cline{3-15} 
\multicolumn{1}{|l|}{}                        & \multicolumn{1}{l|}{}                      & SD       & \multicolumn{1}{l|}{13.3}  & \multicolumn{1}{l|}{14.8}                              & \multicolumn{1}{l|}{16.7}   & \multicolumn{1}{l|}{15.2}                              & \multicolumn{1}{l|}{14.8}   & 10.6  & \multicolumn{1}{l|}{20.3}   & \multicolumn{1}{l|}{20.4}   & \multicolumn{1}{l|}{18.8}   & \multicolumn{1}{l|}{14.4}   & \multicolumn{1}{l|}{12.6}   & 15.3   \\ \cline{2-15} 
\multicolumn{1}{|l|}{}                        & \multicolumn{2}{l|}{Statistic}                        & \multicolumn{1}{l|}{t = 0.91} & \multicolumn{1}{l|}{t = 2.74}                             & \multicolumn{1}{l|}{t = -1.52} & \multicolumn{1}{l|}{t = -1.14}                            & \multicolumn{1}{l|}{t = -0.05} & t = 0.77 & \multicolumn{1}{l|}{t = 0.72}  & \multicolumn{1}{l|}{t = -0.15} & \multicolumn{1}{l|}{t = 1.31}  & \multicolumn{1}{l|}{t = -0.49} & \multicolumn{1}{l|}{t = -0.78} & t = -0.98 \\ \cline{2-15} 
\multicolumn{1}{|l|}{}                        & \multicolumn{2}{l|}{P-Value}                          & \multicolumn{1}{l|}{.377}   & \multicolumn{1}{l|}{\color{orange}$*$} & \multicolumn{1}{l|}{.147}    & \multicolumn{1}{l|}{.268}                               & \multicolumn{1}{l|}{.960}    & .449   & \multicolumn{1}{l|}{.479}    & \multicolumn{1}{l|}{.885}    & \multicolumn{1}{l|}{.208}    & \multicolumn{1}{l|}{.627}    & \multicolumn{1}{l|}{.447}    & .339    \\ \hline
\multicolumn{1}{|l|}{\multirow{6}{*}{Radius}} & \multicolumn{1}{l|}{\multirow{2}{*}{16:9}} & M        & \multicolumn{1}{l|}{1.5}   & \multicolumn{1}{l|}{1.2}                               & \multicolumn{1}{l|}{1.2}    & \multicolumn{1}{l|}{1.1}                               & \multicolumn{1}{l|}{1.1}    & 1.0   & \multicolumn{1}{l|}{1.5}    & \multicolumn{1}{l|}{1.3}    & \multicolumn{1}{l|}{1.3}    & \multicolumn{1}{l|}{1.2}    & \multicolumn{1}{l|}{1.2}    & 1.2    \\ \cline{3-15} 
\multicolumn{1}{|l|}{}                        & \multicolumn{1}{l|}{}                      & SD       & \multicolumn{1}{l|}{0.6}   & \multicolumn{1}{l|}{0.4}                               & \multicolumn{1}{l|}{0.3}    & \multicolumn{1}{l|}{0.2}                               & \multicolumn{1}{l|}{0.3}    & 0.2   & \multicolumn{1}{l|}{0.5}    & \multicolumn{1}{l|}{0.3}    & \multicolumn{1}{l|}{0.4}    & \multicolumn{1}{l|}{0.3}    & \multicolumn{1}{l|}{0.3}    & 0.4    \\ \cline{2-15} 
\multicolumn{1}{|l|}{}                        & \multicolumn{1}{l|}{\multirow{2}{*}{3:2}}  & M        & \multicolumn{1}{l|}{1.4}   & \multicolumn{1}{l|}{1.4}                               & \multicolumn{1}{l|}{1.2}    & \multicolumn{1}{l|}{1.1}                               & \multicolumn{1}{l|}{1.1}    & 1.0   & \multicolumn{1}{l|}{1.6}    & \multicolumn{1}{l|}{1.5}    & \multicolumn{1}{l|}{1.4}    & \multicolumn{1}{l|}{1.2}    & \multicolumn{1}{l|}{1.2}    & 1.3    \\ \cline{3-15} 
\multicolumn{1}{|l|}{}                        & \multicolumn{1}{l|}{}                      & SD       & \multicolumn{1}{l|}{0.5}   & \multicolumn{1}{l|}{0.5}                               & \multicolumn{1}{l|}{0.3}    & \multicolumn{1}{l|}{0.3}                               & \multicolumn{1}{l|}{0.4}    & 0.2   & \multicolumn{1}{l|}{0.7}    & \multicolumn{1}{l|}{0.7}    & \multicolumn{1}{l|}{0.4}    & \multicolumn{1}{l|}{0.4}    & \multicolumn{1}{l|}{0.3}    & 0.5    \\ \cline{2-15} 
\multicolumn{1}{|l|}{}                        & \multicolumn{2}{l|}{Statistic}                        & \multicolumn{1}{l|}{t = 1.50} & \multicolumn{1}{l|}{z = 45.0}                             & \multicolumn{1}{l|}{t = 0.76}  & \multicolumn{1}{l|}{t = 0.25}                             & \multicolumn{1}{l|}{z = 72.0}  & t = 0.60 & \multicolumn{1}{l|}{z = 71.0}  & \multicolumn{1}{l|}{z = 81.0}  & \multicolumn{1}{l|}{t = -0.36} & \multicolumn{1}{l|}{t = -0.39} & \multicolumn{1}{l|}{z = 79.0}  & z = 47.0  \\ \cline{2-15} 
\multicolumn{1}{|l|}{}                        & \multicolumn{2}{l|}{P-Value}                          & \multicolumn{1}{l|}{.151}   & \multicolumn{1}{l|}{.081}                               & \multicolumn{1}{l|}{.457}    & \multicolumn{1}{l|}{.807}                               & \multicolumn{1}{l|}{.580}    & .555   & \multicolumn{1}{l|}{.551}    & \multicolumn{1}{l|}{.865}    & \multicolumn{1}{l|}{.721}    & \multicolumn{1}{l|}{.705}    & \multicolumn{1}{l|}{.799}    & .099    \\ \hline
\end{tabular}
\end{adjustbox}
\end{table*}






%% file: introduction.tex
\IEEEraisesectionheading{\section{Introduction}\label{intro}}
\IEEEPARstart{A}{R} teleconferencing is an essential ingredient of future teleconferencing. It symmetrically overlays agents of spatially separated users on each other's local physical environment, creating {an} 
illusion of human teleportation. With the potential to improve the experience of online interaction to get closer to physically being together, such techniques have emerged as trending topics in both industry and research.

\begin{figure*}
    \centering
    \includegraphics[width=\linewidth]{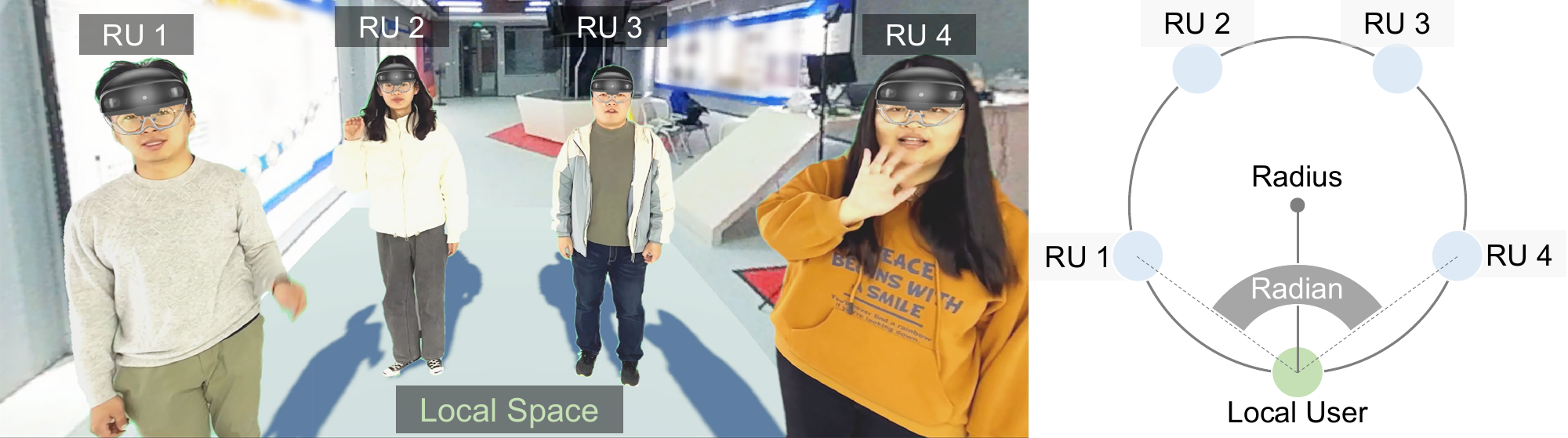} 
    \caption{
    A local user (not shown in this figure) is having a small-group AR teleconference with four remote users through their video avatars. 
    The left scene is seen from the local user's perspective, with the remote users' video avatars (denoted as ``RU 1'' to ``RU 4'') {overlaid} 
    in the local physical space. It is a screenshot from our VR-simulated AR environment {(in the FoV = 110$^\circ$ condition)} 
    with a post-added HoloLens on each RU's head to demonstrate {that every user {has} 
    the conversation through an AR HMD symmetrically} 
    in real-use scenarios.
    The right image illustrates 
    the layout of the conversation group from the top-down view. The local user adjusts the RUs' placement on the circle using {the} 
    Radian and Radius {parameters}.%
    }
    \label{fig:teaser}
\end{figure*}

In pursuit of a high-quality 3D experience, {one group of} existing solutions {present impressive real-time visual-audio effects by} thoroughly building media processing pipelines containing multiple RGB-D cameras, high-bandwidth networking systems, and sophisticatedly designed displays 
\cite{lawrence2021project,orts2016holoportation,zhang2022virtualcube}. However, such high hardware requirements {currently} make these systems too expensive and complicated to set up in daily-use scenarios. 
The prevalence of AR/MR HMDs in recent years brings the opportunity to make the immersive experience more accessible, which is 
why they are considered 
the next-generation personal computing platform in place of smartphones. Therefore, it is critical to find 
the form of teleconferencing applications on AR HMDs that, on the one hand, is mobile, affordable, and easy to access, and on the other hand, can truly provide an upgraded user experience compared to video conferencing {on PCs and mobile devices}. 
To this end, existing solutions directly transplant video conferencing onto the AR HMD platform \cite{MicrosoftRemote,MagicleapAssist} 
or use virtual avatars as the mediation to represent the remote peers \cite{MicrosoftMesh,MetaHorizonWorkrooms,piumsomboon2018mini,yoon2020placement,wang2022predict}.
The direct transplant shows the real appearance of a remote user by placing his/her video on a mid-air virtual screen. It preserves the screen-based video conference's advantage of being photorealistic but fails to immerse the remote users into the local physical environment. Using 3D virtual avatars can create a more immersive experience that is well integrated with the physical environment but significantly loses the fidelity of real {users} and tends to fall into the uncanny valley when increasing the realism of the avatar model. They can match the needs of certain task-oriented usage scenarios such as remote assisting or guidance, but are not enough for interaction-, social-, or user-oriented scenarios \cite{junuzovic2012see,pan2016comparison,de2019watching,yu2021avatars} such as group conversations with friends, formal meetings, and everyday telepresence \cite{yoon2020placement,wang2022predict}, where users want the remote peers to feel both real and present.

To {optimize these two objectives}, 
we {explore the use of} 
{life-size 2D video avatars (simply referred as video avatar{s})} for AR teleconferencing applications on the HMD platform {and examine} 
whether and how they can help to achieve the balance between fidelity and co-presence. We focus on the small-group conversation scenario {studied} 
in prior works \cite{vertegaal1999gaze,billinghurst2000out,otsuka2016mmspace,he2021gazechat} as a fundamental use case in AR teleconferencing. 
There are two types of video-based avatars: namely, 3D video avatars and 2D video avatars, in the literature. 
Volumetric 3D video avatars with both textures and geometries have been used in CAVE or projection-based installations and HMDs in telepresence systems \cite{lee2005toward,pejsa2016room2room,maimone2013general,orts2016holoportation}. 
{A 2D video avatar is a simpler format consisting of only pixels on a plane, which is also explored by previous works \cite{insley1997using,billinghurst2000out,borst2018teacher,cui2022fusing,de2019watching} and is used in our work.} However, they rarely explore the HMD-based AR small-group conversation scenario with multiple remote users as life-size video avatars. {To have immersive interactions with remote users' video avatars, we have to first determine their proper placements regarding users' social proximity.} 
Literature has found certain formations \cite{kendon1990conducting,pathi2019f,marshall2011using} and personal proxemics in both real \cite{hall1968proxemics,hecht2019shape} and virtual \cite{huang2022proxemics,ye2021paval} worlds for one-to-one and one-to-many 
conversations. {{Although} qualitatively predictable, quantifying interpersonal proxemics has been proved important by these works, but is unexplored for video avatars in AR telepresence despite its importance for computationally determining the video avatar placement. Moreover,} with the AR experienced through HMDs being not exactly the ultimate AR indistinguishable from reality, users might perceive the proximity differently due to a potentially undermined immersive experience {caused} by the limited FoV. This leads to our first question: \textbf{what is the user-centered optimal placement for 
life-size video avatar{s} in {an} HMD-based {limit-FoV} AR small group teleconference (RQ1)?} To answer this question, we conduct a pilot study 
{in Sec. \ref{pilot}} to determine the video avatar's 
optimal placement in scenarios with 1, 2, 3, and 4 remote users, regarding the local user's comfortableness and sense of 
co-presence. The results yield references for the video avatar placement 
{in our} subsequent evaluation {in Sec. \ref{Evaluation} and inform the further quantitative study in Sec. \ref{FurtherVRStudy}}.

With the suggested placement from the pilot study, we then implement a proof-of-concept prototype of video-avatar-based AR teleconferencing. Specifically, {it enables remote teleconferencing through video avatar, AR transplant of video conferencing (simply referred as video grid), and avatar representations using a commodity camera and an AR HMD (Microsoft HoloLens 2) for each user. Using our prototype,} 
we compare the video-avatar-based teleconferencing (Video Avatar mode) with conventional video conferencing, direct AR HMD transplant of video conferencing 
(Video Grid mode), and avatar-mediated teleconferencing 
(Avatar mode) in a 2-remote-user scenario to verify \textbf{whether the use of video avatars outperforms other solutions and achieves the balance between fidelity and co-presence (RQ2).} The results show that the video-avatar condition is the most preferred, presenting a sense of co-presence as high as the avatar condition while having a high fidelity similar to the video-conferencing condition.

{While video avatars are proved to be the most feasible solution for current HMD-based AR teleconferencing, their social proxemics perceived by users (obtained from our pilot study) are significantly different from that of real humans. We believe it is due to the limitation of the AR HMD, since currently, it can only} give us a glimpse at the ultimate AR with a huge gap in between, where the limited FoV is one of the most important causes. 
Existing works reveal the negative effect of a small FoV on the performance of {visual} tasks \cite{czerwinski2002women,kishishita2014analysing,ragan2015effects,kim2022display} and a user's perception of virtual agents \cite{lee2018effects,wang2019exploring}. 
This leads to our third question: \textbf{what is the effect of HMD's FoV on the optimal placement of {video avatars} in AR small group teleconferencing (RQ3)?} To answer this question, 
we conduct our further {quantitative} study that involves {a wide range of} FoV conditions (30$^\circ$, 40$^\circ$, 50$^\circ$, 70$^\circ$, 90$^\circ$, and 110$^\circ$) using a VR-simulated AR environment. {It is a common practice to use VR simulation to explore problems from AR with {some} 
hardware constraints \cite{ren2016evaluating,medeiros2022shielding,evangelista2022auit}. We adopt the VR simulation {since} 
VR HMDs have relatively larger FoVs than Optical See-Through (OST) AR HMDs, such as 
HoloLens 2 we used in the pilot study and the evaluation.} 
We also compare the two most common aspect ratios for current AR HMDs (16:9 and 3:2)\footnote{\url{https://vr-compare.com}} in the study design. 
We first {initially verify the validity of the VR simulation by comparing the data from the AR pilot study and the VR study.} 
Then we examine {the effect of} the aspect ratio and find no significant difference in the placement under the two involved conditions. Subsequently, we find significant correlations between FoV and the video avatar placement in each group size, and regress the data into placement models with FoVs ranging from 10$^\circ$ to 180$^\circ$. They can give references for the video avatar placement on not only AR HMDs currently available but also future ones with larger FoVs up to nearly that of human eyes.

The main contributions of this work are threefold.
\begin{itemize}
\item Through a pilot study, we find the optimal user-centered video-avatar placement in small{-}group AR teleconferencing scenarios with {up to} 
4 remote users, and identify the potential influence of the HMD's FoV on the optimal placement.
\item We build a proof-of-concept prototype system using the results obtained from the pilot study. Through a user study {based on this prototype}, we conclude 
the video avatar's ability to achieve a better balance between fidelity and co-presence.
\item We regress placement models from the data collected in a further experiment in a VR-simulated AR environment involving {larger} 
FoV conditions 
to provide {quantitative} references for the video{-}avatar placement in AR teleconferencing applications on current {and future} AR HMDs. 
\end{itemize}



%% file: relatedwork.tex
\section{Related Work}
\subsection{Video-Based Avatars for AR Teleconferencing}
{For 3D video avatars,} the media processing pipeline generally consists of capturing, transmitting, reconstructing, and rendering {the volumetric data}. 
The effect of systems using specially equipped installations to build the pipeline has seen a tremendous evolution from earlier endeavors \cite{prince20023d,lee2005toward,
jones2009achieving,
maimone2013general,beck2013immersive,
pejsa2016room2room} to most recent works \cite{orts2016holoportation,lawrence2021project,zhang2022virtualcube}. To capture and present 3D video avatars, {existing systems rely on complex setups (e.g., fifteen surrounding cameras \cite{prince20023d}, CAVE-like installations  \cite{lee2005toward}, a multi-camera setup and a projection-based display \cite{jones2009achieving}, a set of RGB-D cameras and projectors \cite{maimone2013general,beck2013immersive}, 
two rooms equipped with RGB-D cameras and projectors \cite{pejsa2016room2room}).}
State-of-the-art systems such as Holoportation \cite{orts2016holoportation}, Project Starline from Google \cite{lawrence2021project}, and VirtualCube \cite{zhang2022virtualcube} also develop their own sophisticated installations to demonstrate teleconference experiences with unprecedented fidelity and 3D effect. However, the complicated and expensive configurations currently prevent 
these solutions with high-quality 3D video avatars {from being} 
applied in broader everyday scenarios. 
While using a single consumer-grade RGB-D camera such as 
Microsoft Kinect can also obtain a 3D or ``2.5D'' video avatar \cite{gamelin2021point,yu2021avatars}, it still suffers from artifacts with the effect more similar to that of {those} methods at the earlier stage.

While it is hard for 3D video avatars to be both high-fidelity and easy to capture and reconstruct, real-time video streaming makes it very simple for 2D video avatars. It has been decades for researchers to incorporate 2D human videos into VR \cite{insley1997using,yura1999video,wang2007contextualized,borst2018teacher,prins2018togethervr,cui2022fusing} and AR \cite{billinghurst1999real,billinghurst2000out,de2019watching} environments. 
{For example, }Insley et al. \cite{insley1997using} pre-record humans from every angle into videos and select {an} 
appropriate image frame according to the relative position of {a} 
viewer to the video avatar to display in a CAVE VR environment. Yura et al. \cite{yura1999video} embed a 2D video avatar into a collaborative virtual environment and change the video stream according to the perspective between the viewer and the video avatar as well. Prins et al. \cite{prins2018togethervr} use video avatars to represent remote users in social VR. Borst et al. \cite{borst2018teacher} incorporate a teacher as a video avatar in VR education scenarios. Although they capture the teacher using an RGB-D camera and render him as a 3D mesh, showing majorly the frontal view with little movement leads to an effect with fewer artifacts and is similar to {a} 
2D video avatar. Cui et al. \cite{cui2022fusing}, and Wang et al. \cite{wang2007contextualized} fuse video avatars into virtual scene models for surveillance use. Billinghurst et al. \cite{billinghurst1999real,billinghurst2000out} overlay videos of users in the real world aligned with the corresponding objects through AR HMD using marker-based tracking. Simone et al. \cite{de2019watching} study the use of a video avatar for watching TV together in a VR-simulated AR environment. They {either do not use life-size portraits or} concentrate on the system implementation and how to display the 2D video avatar from the right perspective to make the experience more 3D-like. 
We focus more on the {life-size whole-body} video avatar's placement, its performance regarding fidelity and presence, and how the placement is affected by the HMD's FoV in the user-centered group conversation scenario.

Telepresence robots such as Double\footnote{\url{https://www.doublerobotics.com}} represent another form of using 2D video avatars in teleconference and telepresence scenarios. Such approaches mount a screen on a movable robot controlled by a remote user as his/her local representation. Otsuka \cite{otsuka2016mmspace} uses actuated screens to present small group-to-group teleconferences. Neustaedter et al. \cite{neustaedter2018from} explore the usage of telepresence robots to attend academic meetings. Furuya and Takashio \cite{furuya2020telepresence} propose to improve the user experience of telepresence robots by replacing the background of the remote user's video on the screen with the local environment, blending the robot more into the surroundings. While being a practical solution, the form of a robot with the extra hardware component makes it less similar to real humans than placing life-size 
video avatars in AR. Moreover, the AR solution is more scalable to multi-user scenarios.

We conclude that compared to existing solutions, using life-size 2D video avatars in {small-group} HMD-based AR 
teleconferencing is more feasible with its low cost and easy setup, and the ability to present a rather high-fidelity experience with fewer artifacts in this user-centered layout.

\subsection{Content Placement in AR}

Existing works exploring the placement of virtual agents in AR and VR mostly focus on the scene semantics. {They statically place them on a sofa or a seat \cite{maimone2013general,pejsa2016room2room} or retarget them according to the semantics of the remote user's position \cite{jo2015spacetime,yoon2020placement}.} For the video avatar placement in our scenario, the main considerations are the user's comfortableness and perceived co-presence and similarity to real-human scenarios, and the potential influence of the HMD's FoV. {{The works by} Lang et al. \cite{lang2019virtual} and Ye et al. \cite{ye2021paval} share a similar goal to ours by considering the agent-user interaction. But they focus on virtual avatars in assistant and navigation scenarios while we explore video avatars in the AR teleconferencing scenario.}

From the perspective of User Interface (UI), researchers also explore the placement of UI components in AR and VR regarding user's social and physical comfortableness in public transit \cite{medeiros2022shielding}, the UI elements' relation{s} with each other and the physical environment \cite{evangelista2022auit}, and users’ cognitive load \cite{lindlbauer2019context}. With the video avatar being a special type of UI element, besides ergonomics, we also focus on interpersonal relations \cite{bailenson2003interpersonal} and the user's feel of proxemics \cite{huang2022proxemics} when determining the placement of {remote users' video avatars}. 

\subsection{Limited Field of View in AR}
One of the major restrictions of the current AR HMD is the limited FoV. Being not able to cover the full FoV of human eyes (around 200$^\circ$ horizontally and 150$^\circ$ vertically), the small FoV of HMDs can cause problems for users during the immersive experience. On the one hand, it can harm task performance. Czerwinski et al. \cite{czerwinski2002women} find it slows the navigation speed in virtual environments. Ragan et al. \cite{ragan2015effects} find it worsens training effectiveness for a visual scanning task. Kim et al. \cite{kim2022display} find it increases the targeting time in gaming. Ren et al. \cite{ren2016evaluating} find it increases the task completion time in an information-seeking scenario. On the other hand, it can affect the user's perception of virtual content. Lee et al. \cite{lee2018effects} find the mismatch between the FoVs of the HMD and human eyes affects the user's behavior related to the perception of the virtual agent in the proximity. Wang et al. \cite{wang2019exploring} discuss the potential impact of small FoV on users' sense of presence or the agent's realism. Swan et al. \cite{swan2007egocentric} find the existence of an occluder (resulting in a decreased FoV) leads to more misjudgments on distance.
With these previously-found various effects caused by FoV, we explore whether it can affect the optimal placement of video avatars in our group conversation scenario.

\subsection{Perception of Remote Participant Representation}
{Researchers find video-based representations to be more preferred than virtual avatars in social-oriented scenarios involving more user-agent interactions such as conferencing for work \cite{junuzovic2012see}, advice seeking \cite{pan2016comparison}, teleconsultation \cite{yu2021avatars}, education \cite{woodworth2019evaluating}, and TV-watching \cite{de2019watching}}. 
Other than the video avatar perceived from the third-person perspective, existing works show an improved experience with self-embodied video avatars. Gonzalez et al. \cite{gonzalez2022bringing} find using a video avatar 
to present the self embodiment can lead to a high level of sense of presence and embodiment, and find it provides better visual quality than using virtual replicas \cite{morin2022full}. Lee et al. \cite{lee2016enhancing} find incorporating the self-embodied video avatar can enhance the AR cinematic experience. {Similar positive effects 
of video-based representation {were} found on size estimation, body ownership, and presence \cite{jung2018over}.}

{These works motivated us to further investigate the use of video avatars in teleconferencing.} Our work explores the user's perception by comparing the video avatar representation with screen-based video conferencing, direct transplant of video conferencing, and avatar-mediated teleconferencing. We {find video avatars} 
can provide a high level of both fidelity and co-presence in this user-centered layout, with an experience surpassing those of the direct transplant and the virtual avatar-mediated approach in such teleconferencing applications on AR HMDs.







%% file: pilotstudy.tex
\section{Pilot Study\label{pilot}}

In this section, we introduce our pilot study on the optimal placement of the remote participants' video avatars in the local space. As mentioned previously in this paper, users often place themselves following certain formations in real-world group conversations and tend to have different perceptions of proxemics and co-presence in the virtual world with the change in factors such as remote user representation and FoV. Therefore, with life-size 2D video avatars 
in the small-group conversation scenario unexplored, we conduct our pilot study to find their optimal placements that serve the best for group conversations with different numbers of participants, and initially explore the effect of FoV on the optimal placement {on the widely adopted Microsoft HoloLens 2 OST AR HMD}.

\subsection{Participants}
We recruited 18 graduate students (14 males and 4 females; average age: 23.6 (SD = 0.7)) from the local campus as the participants. 
{The sample size (along with that in the subsequent user study in Sec. \ref{Evaluation}) is in line with the suggestion by relevant research in HCI studies~\cite{caine2016local}, and is thus believed to be sufficient for drawing our conclusions. We recruited young college students as participants since they are sensitive to XR technology and are our main target users. {They can represent the population who are willing to try and judge new AR/VR technologies.}} 
{It is also a common sample composition in other AR/VR studies \cite{woodworth2019evaluating,yoon2020placement,kim2021adjusting}. We mainly focused on the diversity of the participant's AR/VR experience since the novelty effect could influence the result of studies on new technologies. Among the participants,} 
5 had no prior AR/VR experience, 9 had experienced several times, and 4 had used HMDs extensively.

\subsection{Study Scenarios \label{ScenariosAR}}
We conduct our pilot study in the {small-group} symmetric 
AR 
teleconference scenario. In this scenario, users are {located} 
in different spaces and join the teleconference using their 
AR HMDs as video avatars. {Each user has the same setup and an equal role} 
with no superior characters such as leaders, teachers, or interviewers. The relation between each user is acquaintances or friends, with no extra intimacy or hostility. The conversation topics are free. People can talk about serious topics such as work and research {or} 
casual topics such as dinner suggestions and games. We set four scenarios with the difference only in the 
conversation group {size}. We denote {them} 
as 1-RU, 2-RU, 3-RU, and 4-RU scenarios, including 1, 2, 3, and 4 remote users (a total of 2, 3, 4, and 5 participants{, with the local user included}), 
respectively.

\subsection{Experiment Setup}
As mentioned earlier, we use Microsoft HoloLens 2 to present the AR experience in our pilot study. It has a diagonal FoV of approximately 50$^\circ$ 
with the 3:2 aspect ratio. We pre-record the videos of every remote user greeting the local user and each other in all four scenarios, remove the background using MobileNetV3 \cite{howard2019searching}, and place them in Unity for HoloLens to render. To ensure the video avatars are life-size, it is 
important to calibrate their sizes to be the same as the corresponding users. So we let {each} user being recorded in the corresponding video clip stand next to his/her video avatar (observed through HoloLens) as the reference for the calibration to make sure they are of the same height and width.

Informed by previous studies on the F-Formation in group conversations \cite{marshall2011using} that people commonly tend to form a circle in conversations with multiple people, we set a circular layout for our AR small-group conversation scenario, as shown in Fig. \ref{fig:teaser} {(Right)}. We pin the local user at the bottom of the circle and symmetrically place the RUs on the circle. The final placement is determined by two parameters: Radius and Radian. Radius defines the size of the circle. It is also perceived as the vertical distance from the local user to the group of remote users' video avatars. A large Radius puts the video avatars far away from the local user, while a small Radius pull them near. Radian defines the angle range from the video avatar on the leftmost to the one on the rightmost. Since the video avatars distribute {evenly in} 
the Radian range, it is also perceived as the horizontal distance between {adjacent video avatars}. 
A large Radian separates the video avatars {from each other} and makes the ones on the leftmost and rightmost get closer to the local user, while a small Radian gathers the video avatars towards the top of the circle. We let the participants adjust the video avatars' placement using {`IJKL' keys on a keyboard}.


We control the FoV by placing four 
black virtual occluders $0.3m$ away in front of the main camera and manipulating their positions accordingly in each FoV condition. With pure black being transparent in OST HMDs, {the occluders shape the FoV by clipping certain parts of the virtual content on the top, bottom, left, and right}, leaving the physical space unaltered.

\begin{figure}[tbp]
  \centering
  \includegraphics[width=\linewidth]{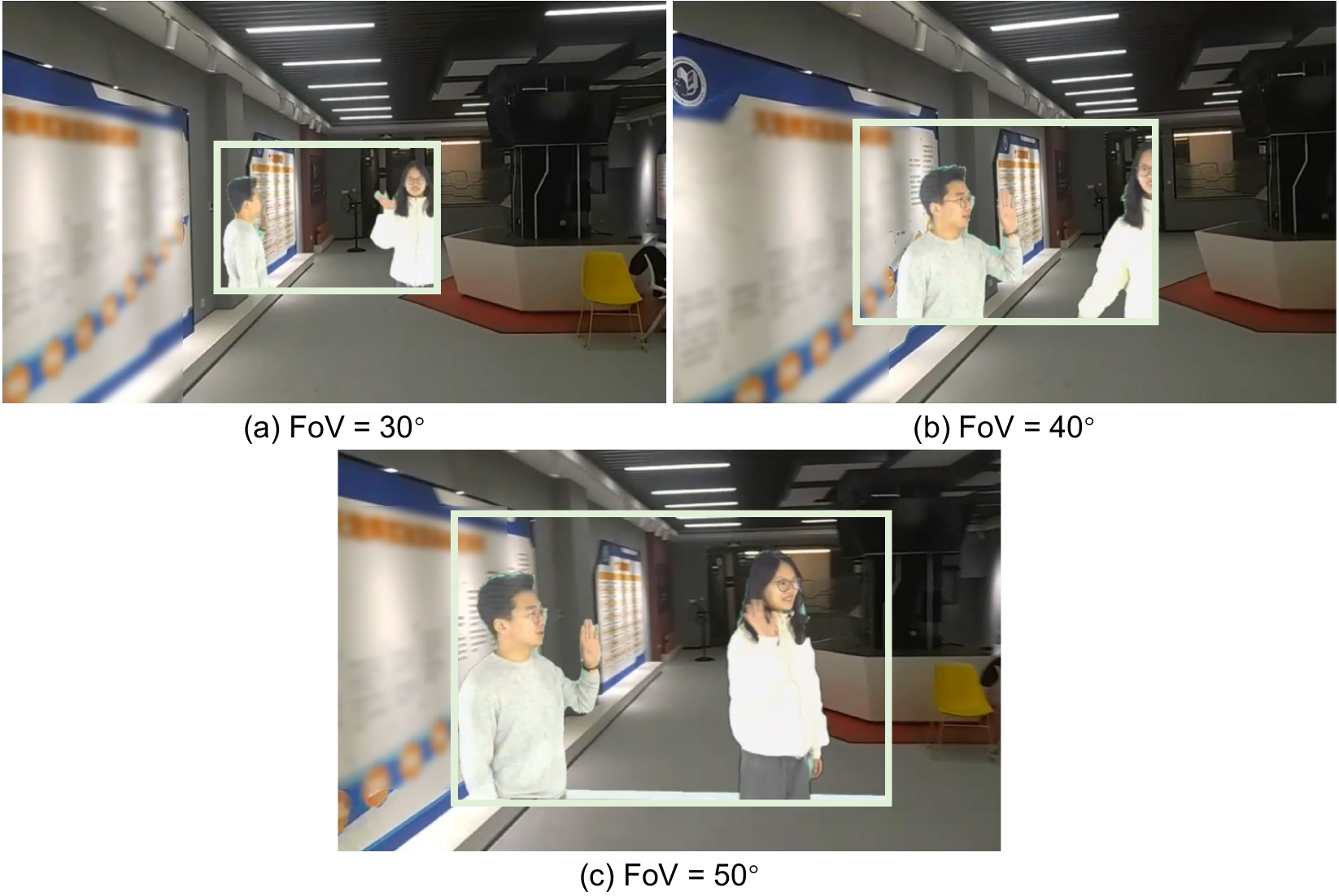}
  \caption{{An illustration of the (a) 30-, (b) 40-, and (c) 50-FoV conditions in the pilot study.} 
  Here in the {images}, 
  {boundaries of the FoV} {(i.e., rectangles in white)} are approximately annotated and are not visible to the participant. {The virtual content outside of the FoV is clipped by the transparent occluders, whose }exact positions 
  for each FoV condition are calculated and defined in the controller script. {Each image shows the view after users adjust the video avatars’ placement under the corresponding FoV condition. The change in the user's perceived optimal $Radius$ leads to different sizes of video avatars in the images.}
  }
  \label{fig:ARFoVConditions}
\end{figure}

We consider three FoV conditions, i.e., \textbf{30-FoV} (FoV = 30$^\circ$), \textbf{40-FoV} (FoV = 40$^\circ$), and \textbf{50-FoV} (FoV = 50$^\circ$). {They cover the FoV range of common current-available OST HMDs. Here in the pilot study, we use FoV as a simple reference to the diagonal FoV with an aspect ratio of 3:2.} We annotate the coarse FoV boundaries under the three FoV conditions in Fig. \ref{fig:ARFoVConditions}. The {images} 
are screenshots from a participant's trial (\textit{p5}) in the 3-RU scenario when the video avatars are placed at the distance the participant felt optimal.


\subsection{Metrics \label{MetricsAR}}
We ask the participants to {decide the optimal distance by considering three metrics}, 
namely comfortableness, sense of co-presence, and similarity to the real-world scenario. A high level of comfortableness means the participant would not feel too much fatigue during the conversation and feel everyone's placement is socially appropriate. According to previous measurements \cite{nowak2003effect}, we describe the sense of co-presence to {the} participants as the feeling that the video avatars ``create 
a sense of distance between you'', ``they actually take up a spot in the physical space'', and ``you have the urge to spatially interact with them''. High similarity to the real-world scenario means {each video avatar's placement} 
in the AR experience {is} 
very similar to {that} 
in a real-world conversation where the remote users are physically co-located in the space. We ask the participants to find the video avatars' optimal placement to {balance} 
the three objectives of optimizing the three corresponding metrics.

\subsection{Procedure}
First, we introduce the participants the basic concept of AR teleconferencing. Second, after they get {its idea}, 
we introduce the small-group conversation scenario in our study (as introduced in Sec. \ref{ScenariosAR}) verbally using real-life examples and tell them their role {as} 
the local user. {Afterward}, 
we introduce the participants {the} 
task of finding the optimal placement for the remote users' video avatars. We introduce the meaning of the three metrics to {them} 
using the definition in Sec. \ref{MetricsAR} and give examples for their better understanding. We {ensure} 
they understand each metric and clarify the optimal placement being the result of balancing all three metrics. Afterward, we introduce how they can adjust the video avatars' placement and submit the result in each task using the keyboard, and walk them through a practice task to get familiar with the operation and the experience. After completing the preparatory work, we proceed to the formal experiment process.

{The experiment has} 
12 tasks in total (3 FoV conditions in each of the 4 scenarios). The sequence of the 3 FoV conditions (corresponding to the 3 tasks in each scenario) is counterbalanced hierarchically using a balanced Latin square. In each task, the participant submits the optimal placement after the adjustment, and the controller script saves the $(Radius, Radian)$ parameter pair. After finishing all 12 tasks, we have a short {semi-structured} interview {with} each participant {that lasts about five minutes} to explore potential insights. {In the interview, we mainly discuss how {they feel} 
during the experiment, what {they think} 
is the major reason for the described experience, and suggestions on how to improve the proposed system.}


\begin{figure*}[htb]
  \centering
  \includegraphics[width=\textwidth]{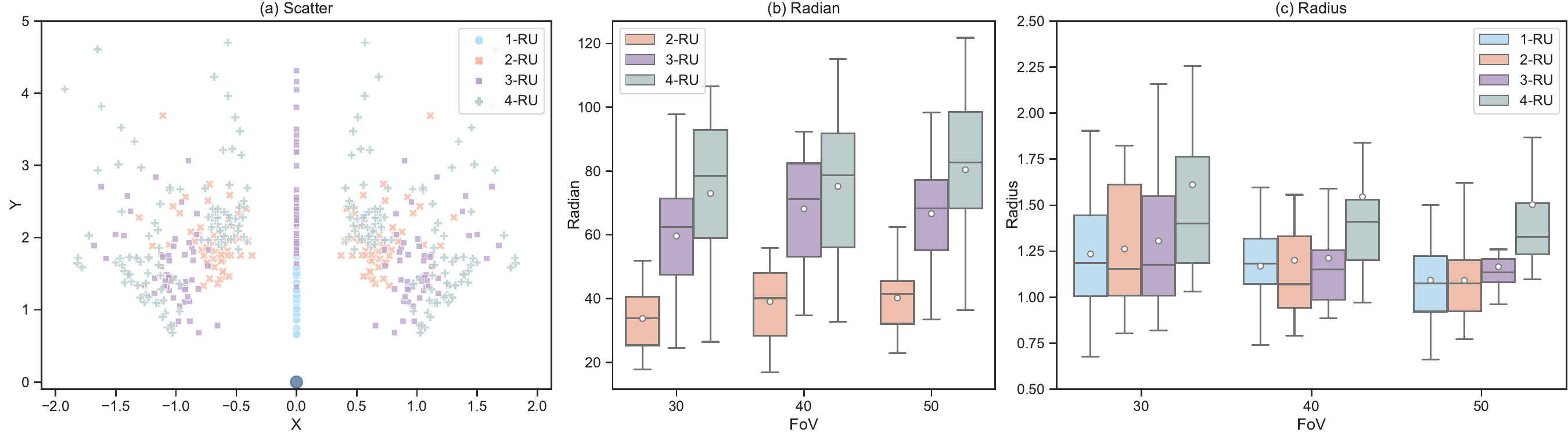}
  \caption{%
  	The statistical results for the data of the pilot study. (a) shows the distribution of all the placement data in all four scenarios (shown in different colors and styles). The solid point in dark blue at 
   (0, 0) 
   denotes the local user. (b) shows the comparison of the Radian data in 2-RU, 3-RU, and 4-RU scenarios since the participant stands directly face-to-face with the remote peer with a radian = $0$ in the 1-RU scenario. (c) shows the comparison of the Radius data in all four scenarios but 
   {with outliers hidden} 
   since they contribute little to the overall trend and some of them are too far away, as 
   seen in (a).
  }
  \label{fig:ARRUPlacement}
\end{figure*}

\subsection{Results}
We collected $18\times3\times4 = 216$ sets of $(Radius, Radian)$ data in total (Radian in the 1-RU scenario has no meaning). Fig. \ref{fig:ARRUPlacement} shows the distribution and statistical results. 
{The results in each condition helps determine the quantitative proxemics for placing teleconferencing video avatars on OST HMDs with corresponding FoVs. The comparison of the results from the three conditions can help initially explore the FoV's effects on the video avatar placement in this small range.} To analyze the results, {following the analysis in similar works \cite{yoon2020placement,ye2021paval},} we first ran Shapiro-Wilk Normality Tests on the collected data to {check} their normality. We then conducted pairwise T-Tests (if the sample is normally distributed, otherwise, non-parametric {Wilcoxon Signed-Rank Tests}) 
on the data under \textbf{30-FoV} 
and \textbf{50-FoV} conditions for each scenario to see if there are significant differences in the $Radian$ and $Radius$ values between them. {We examined the difference between {\textbf{30-FoV} and \textbf{50-FoV}} 
because the range of the FoV is rather limited and we want to acquire a preliminary result from a range as long as possible.} We also conducted Pearson Correlation Tests on the normally distributed samples and Spearman Correlation Tests on the non-normally distributed samples to see if the $Radian$ and $Radius$ values significantly correlate with the $FoV$ values. We will elaborate on the specific results under the confidence interval of $p < .05$ below.

\emph{Optimal placement.} For the optimal placement, we distill the average $Radian$ and $Radius$ in each scenario as the parameter pair {to determine} 
the optimal placement. Note that the Radian is measured in degrees, and the Radius is measured in meters. {We list the data for the \textbf{30-} and \textbf{40-FoV} conditions in the supplemental file (Sec. 2.1), and report the data in the \textbf{50-FoV} condition below, which will be used} in the subsequent evaluation in Sec. \ref{Evaluation}. 
For the \textbf{50-FoV} condition, the optimal $(Radian, Radius)$ is $(/, 1.09)$ for the 1-RU scenario, $(40.20, 1.09)$ for the 2-RU scenario, $(66.63, 1.17)$ for the 3-RU scenario, and $(80.42, 1.50)$ for the 4-RU scenario. 

\emph{Initial analysis of the FoV's effect.} 
We find no significant correlation between $(Radian, Radius)$ and FoV. We list the detailed statistical results of the correlation test in Tab. 1 in the supplemental file. We report the data below for the significant test between \textbf{30-FoV} and \textbf{50-FoV} conditions {($N = 18$ in each scenario)}. 

\textit{Radian}. The results show that $Radian$s are significantly smaller in the \textbf{30-FoV} condition 
($M = 33.8$, $SD = 10.4$ in 2-RU scenario, 
$M = 59.6$, $SD = 18.3$ in 3-RU scenario, and 
$M = 73.0$, $SD = 23.6$ in 4-RU scenario) 
than in the \textbf{50-FoV} condition 
($M = 40.2$, $SD = 10.4$ in 2-RU scenario, 
$M = 66.6$, $SD = 15.0$ in 3-RU scenario, and 
$M = 80.4$, $SD = 23.1$ in 4-RU scenario) 
in all three scenarios. 
($t$-$statistic = -4.85$, $p < .001$, in 2-RU scenario, 
$t$-$statistic = -3.22$, $p = .005$ in 3-RU Condition, and 
$t$-$statistic = -3.08$, $p = .007$ in 4-RU scenario). 

\textit{Radius}. The results show that there are no significant differences in $Radius$ between \textbf{30-FoV} 
($M = 1.2$, $SD = 0.3$ in 1-RU scenario, 
$M = 1.3$, $SD = 0.3$ in 2-RU scenario, 
$M = 1.3$, $SD = 0.4$ in 3-RU scenario, and 
$M = 1.6$, $SD = 0.7$ in 4-RU scenario) and \textbf{50-FoV} 
($M = 1.1$, $SD = 0.3$ in 1-RU scenario, 
$M = 1.1$, $SD = 0.2$ in 2-RU scenario, 
$M = 1.2$, $SD = 0.2$ in 3-RU scenario, and 
$M = 1.5$, $SD = 0.5$ in 4-RU scenario) conditions in {1-, 3-, and 4-RU} scenarios. 
($t$-$statistic = 2.01$, $p = .060$ in 1-RU scenario, 
{$z = 47.0$, $p = .099$ in the 3-RU scenario, and 
$z = 61.0$, $p = .304$ in the 4-RU scenario). $Radius$ is significantly smaller in \textbf{50-FoV} in the 2-RU scenario, $z = 38.0$, $p = .038$}.

\subsection{Study Conclusion and Discussion}
From the pilot study, we obtain the video avatars' optimal placement in four scenarios on Microsoft HoloLens 2. We distill the results for the subsequent evaluation {(Sec. \ref{Evaluation})} using the same HMD {and for other OST HMDs with FoVs corresponding to the study conditions}. We also initially analyzed the potential effect of FoV on the optimal placement, which informs the subsequent further 
study (Sec. \ref{FurtherVRStudy}) 
to {quantitatively} examine the effect with a larger range of FoV. Next, we discuss some of the observations in detail.

\textbf{FoV can partially affect the optimal placement of the remote users, especially horizontally.} The significantly smaller $Radian$ in the \textbf{30-FoV} condition than {that} in the \textbf{50-FoV} condition tells that local users tend to let the remote users gather closer to each other to make sure they stay in the area where he/she can be aware of their existence {without rotating their heads too often}. {However, the local user would not simply place the remote users' video avatars tightly next to each other to fit the FoV window because ``it would feel like I am interviewing them and they are closer to each other than I am.'' and ``they would feel not very present when stuck tightly to each other''. For $Radius$,} some participants reflect that ``I {cannot} see the whole body of the remote peers, so I have to let them stand further away from me during the whole process to see at least the upper body, but not too far to {feel} 
unnatural for the group conversation''. We think the main reason for the insignificant effect of $FoV$ on $Radius$ {in most scenarios} is that the $FoV$ conditions are all too small to observe the remote user's whole body without vertical head rotations. Users have to put the video avatars to the upper limit of the distance for natural conversations in all three $FoV$ conditions for a better visual experience while preserving an acceptable spatial relation.

\textbf{Insignificant correlation.} We think it is because the $FoV$ values in the three conditions are all too small to show the difference, and the number of $FoV$ conditions is too limited to reveal the correlation. We can observe changes in $Means$ between different $FoV$ conditions in the data report and in Fig. \ref{fig:ARRUPlacement} (b) and (c) for $Radian$ and $Radius$ respectively. Though not significantly, there are tendencies that $Radian$ would increase and $Radius$ would decrease when $FoV$ becomes larger. These indications are part of the motivation for our Study 3 {in Sec. \ref{FurtherVRStudy}} to include more and larger $FoV$ conditions to find out the correlation.

%% file: evaluation.tex
\section{Implementation and Evaluation\label{Evaluation}}
As elaborated previously in this paper, using video avatars helps make HMD-based AR teleconferencing easy to set up and has the potential to achieve a balance between fidelity and co-presence. Therefore, we first implement a proof-of-concept prototype system to demonstrate the use of video avatars. Then we use the prototype as the platform to conduct an evaluation to verify that using video avatars can present the aforementioned balanced experience.

\subsection{Proof-of-Concept Implementation}\label{SecProofOfConcept}
\begin{figure}[tbp]
  \centering
  \includegraphics[width=\linewidth]{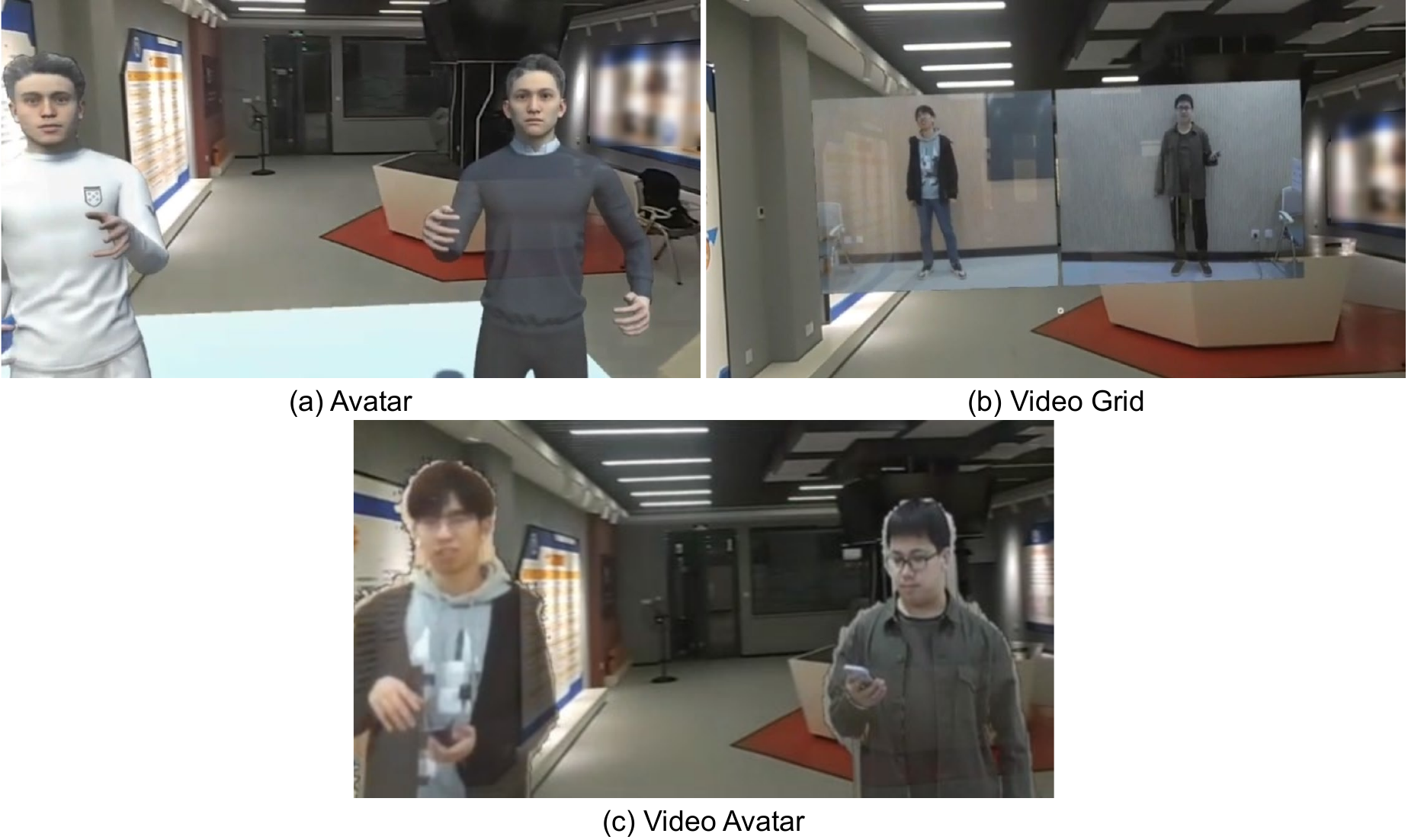}
  \caption{{An illustration} 
  of the (a) Avatar, (b) Video grid, and (c) Video avatar conditions in the evaluation using our prototype system.
  }
  \label{fig:EvaluationConditions}
\end{figure}

Our prototype has three modes (as illustrated in Fig. \ref{fig:EvaluationConditions}), 
namely Avatar mode, Video Grid mode, and Video Avatar mode, where remote users are represented as avatars, videos on AR virtual screens (the direct transplant of video conferencing), and video avatars, respectively. For each user, our system setup includes only an AR HMD to render 
the remote users' agents and a PC with a webcam to capture and process the real-time image of the local user.

In the Avatar mode, the PC runs a 3D human pose estimation script using the video captured by the webcam, and transmits the estimated joint position to other users' HMDs to get the corresponding avatar moving. We use the solution from MediaPipe\footnote{\url{https://mediapipe.dev}} to implement the pose estimation. We build TCP connections for data transmission. Referencing the aforementioned avatar-mediated teleconference applications such as Microsoft Mesh, we only enable the avatar's hand movement, with legs staying in an Idle animation state. We use the positions of the user's wrist joints and apply Inverse Kinematics to present the avatar's arm motion. For the head motion, we make the avatar look at the local user during the whole experience. The avatar and the animation clips in our implementation are from Mixamo\footnote{\url{https://www.mixamo.com}}.

In the Video Grid mode, the PC directly transmits the video frames of the local user to other users' HMDs through TCP networking. The videos are rendered on a plane as a virtual screen placed in front of the local user at {the} head level.

In the Video Avatar mode, the PC runs a background removal script to separate the human from the background in the video captured by the webcam and transmits the matted frames to other users' HMDs through TCP networking as video avatars. The bandwidth requirement in our prototype is only around 300Kb/s in the Video Grid and Video Avatar modes, and lower than 10Kb/s in the Avatar mode since we only transmit the joint positions. 

{We manually adjust virtual light sources to match the physical lighting condition. We also set a virtual floor with the same texture as the physical floor in the room. This setup makes the lighting and shadowing effect of virtual content similar to that of physical objects in the AR environment. This process can be further automated following relative research \cite{rhee2017mr360}.}


\subsection{Participants}
We recruited 20 graduate students (15 males and 5 females; average age: 23.3 (SD = 1.2)) from the local campus as the participants. Most of them had a moderate level of AR/VR experience: 8 had no prior AR/VR experience, 11 had experienced several times, and 1 had used HMDs extensively.

\subsection{Study Scenario}
The scenario in the evaluation is still the small-group conversation 
introduced in the pilot study in Sec. \ref{ScenariosAR}. {We} let the participants play a debate-like conversational game called ``who is the spy'' {with two remote users}. It is utilized in previous studies \cite{he2021lookatchat} as the task for {evaluating} 
group conversation applications. 
In this game, all users get a word of their own, with one of them being different (the spy). Users describe their words in a round and take a vote on who is {the suspected spy}. 
The spy wins if not voted out, and others win if they correctly spot the spy.

\subsection{Experiment Setup}
We use HoloLens 2 to present the AR experience using our prototype system and a commercial application 
to present the conventional video conferencing experience. We have two researchers to play the role of two remote users in the game {and the participant joins as the local user, forming a three-person group conversation for each participant.} We set the remote users' video avatars at the optimal position obtained from our pilot study. We pre-calibrate the distance between the remote users and their webcams to make sure their video avatars observed by the participant are life-size.

\subsection{Study Conditions}
We set four conditions, namely \textbf{Video}, \textbf{Avatar}, \textbf{Video Grid}, and \textbf{Video Avatar}. \textbf{Video} is the conventional video conferencing condition. Users complete the task on the PC with remote users' videos on the screen. The \textbf{Avatar}, \textbf{Video Grid}, and \textbf{Video Avatar} conditions correspond to the three modes of our prototype by name, as illustrated in Sec. \ref{SecProofOfConcept}. {We set the \textbf{Avatar} condition to represent the effect of most prevalent commercial apps such as Microsoft Mesh and Meta Horizon Workrooms. They use simplified cartoon avatars instead of life-like avatars to avoid being uncanny due to the lack of detailed real appearance and animation. In this evaluation, the conditions are within the scope of being mobile and easy to set up. The \textbf{Avatar} and \textbf{Video Grid} conditions represent the effect of existing HMD-based solutions. 
Complicated solutions such as high-quality full-body volumetric capturing and reconstruction are thus not considered.} 
Since the effect of conventional video conferencing is well-known, we show the three AR conditions in Fig. \ref{fig:EvaluationConditions}. 

\subsection{Metrics\label{metricsEvaluation}}
{By referring} 
to similar studies \cite{junuzovic2012see,yu2021avatars}{, we} 
identify three metrics, namely ``Fidelity/Realism'', ``Co-Presence'', and ``Preference''. The ``Fidelity/Realism'' is defined as how photorealistic the remote user's representation is. We also describe it to the participants as ``to what extent you think they are real people or can recognize them''. ``Co-Presence'' is defined and described the same as in the pilot study in Sec. \ref{MetricsAR}. ``Preference'' is an overall subjective rating of the experience. We present the three metrics to the participants using a seven-point Likert scale ranging from 1 to 7. 1 and 7 represent the lowest and highest ``Fidelity/Realism'', ``Co-Presence'', and ``Preference'', respectively.

\subsection{Procedure}
We first introduce the participant{s} the task and make sure they are familiar with the rules of the game. Then we explain that they would use either a PC or an AR HMD to do the task in different conditions. Afterward, we introduce the definition of each metric and {clarify them} 
by giving examples and descriptions. After giving instructions on how to use the AR HMD, we {start} 
the experiment.

In the experiment, the participants 
complete four tasks, one for each condition{, in which the representation of remote users (the two researchers) changes accordingly}. The sequence of the tasks is counterbalanced {among the} 
participants using a balanced Latin square. We remind the participant{s} of the metrics' definitions and the meanings of the scores (as in Sec. \ref{metricsEvaluation}) and ask them to rate their experience by selecting scores under the title of each metric on the seven-point Likert scale introduced earlier after each task. They can change the score of any task at any time. After they finish all four tasks, we {have a five-minute semi-structured interview for them similar to that in the pilot study} 
to explore the potential insights and suggestions.

We make the following hypotheses:
\begin{itemize}
\item \textbf{H1}: The Fidelity of the \textbf{Avatar} condition will be significantly lower than {the} 
video-based conditions.
\item \textbf{H2}: The Co-Presence of \textbf{Avatar} and \textbf{Video Avatar} conditions will be significantly higher than \textbf{Video} and \textbf{Video Grid}.
\item \textbf{H3}: {Compared to the traditional \textbf{Video} condition,} users will prefer the AR-HMD-based conditions more {and prefer} 
\textbf{Video Avatar} {the most}. 
\end{itemize}

\subsection{Results}
\begin{figure}[tb]
  \centering
  \includegraphics[width=0.8\columnwidth]{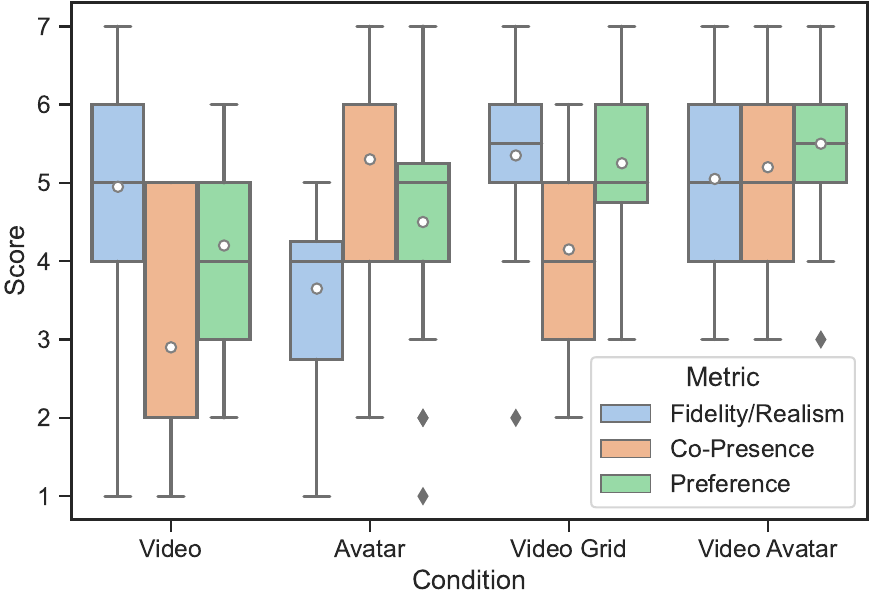}
  \caption{%
  	The statistical results for the data of the user study in the Evaluation.
  }
  \label{fig:EvaluationResult}
\end{figure}

To analyze the study results, we ran Shapiro-Wilk normality tests on the collected data and found non-normally distributed samples. Therefore, we conducted non-parametric {Friedman Test and Post-hoc Nemenyi Test} 
to compare the subjective ratings of all the metrics. We will elaborate on the specific results under the confidence interval of $p < .05$ below. The statistical results for the data are shown in Fig. \ref{fig:EvaluationResult}.

\emph{Fidelity/Realism.}
The {Friedman Test} shows a significant effect of the four conditions
({$Chi$-$square$  $statistic = 18.66$, $p < .001$}). We further conducted {{the} Post-hoc Nemenyi Test} for pairwise comparisons and summarized the results in Tab. \ref{tab:Fidelity/Realism}. The fidelity of \textbf{Avatar} ($M = 3.7$, $SD = 1.2$) is significantly lower than \textbf{Video} ($M = 5.0$, $SD = 1.6$), \textbf{Video Grid} ($M = 5.4$, $SD = 1.1$), and \textbf{Video Avatar} ($M = 5.1$, $SD = 1.0$). There is no significant difference between each two of the three video-based representations (i.e., \textbf{Video}, \textbf{Video Grid}, and \textbf{Video Avatar}).

\begin{table}[tb]
  \caption{{Post-hoc Nemenyi Test results for Fidelity/Realism in the Evaluation. ($*$ $p < .05$, $**$ $p < .01$, $***$ $p < .001$)}}
  \label{tab:Fidelity/Realism}
  \scriptsize
  \centering
  \begin{tabu}{r*{7}{c}*{2}{r}}
  	\toprule
  	\textbf{Fidelity/Realism} &  Video & Avatar & Video Grid & Video Avatar\\
  	\midrule
  	Video       &  $/$   &       &        &     \\
  	Avatar      &  $*$   &  $/$  &        &     \\
  	Video Grid   & $.900$& $**$  &   $/$  &     \\
  	Video Avatar & $.900$& $*$   & $.900$ & $/$ \\
  	\bottomrule
  \end{tabu}
\end{table}

\emph{Co-presence.}
The {Friedman Test} shows a significant effect of the four conditions
({$Chi$-$square$  $statistic = 26.93$, $p < .001$}). We further conducted {{the} Post-hoc Nemenyi Test} for pairwise comparisons and summarized the results in Tab. \ref{tab:Co-Presence}. We can see that 
{the co-presence score of \textbf{Avatar} ($M = 5.3$, $SD = 1.4$) is significantly higher than that of \textbf{Video} ($M = 2.9$, $SD = 1.5$), but is not significantly different from that of \textbf{Video Grid} ($M = 4.2$, $SD = 1.3$). \textbf{Video Avatar} ($M = 5.2$, $SD = 1.3$) can achieve a significantly higher level of co-presence than both \textbf{Video} and \textbf{Video Grid}, reaching a high level with no significant difference with \textbf{Avatar}. There is no significant difference between \textbf{Video Grid} and \textbf{Video} regarding co-presence.}

\begin{table}[tb]
  \caption{{Post-hoc Nemenyi Test results for Co-Presence in the Evaluation. ($*$ $p < .05$, $**$ $p < .01$, $***$ $p < .001$)}}
  \label{tab:Co-Presence}
  \scriptsize
  \centering
  \begin{tabu}{r*{7}{c}*{2}{r}}
  	\toprule
  	\textbf{Co-Presence} &  Video & Avatar & Video Grid & Video Avatar\\
  	\midrule
  	Video       &  $/$   &       &        &     \\
  	Avatar      &$***$   &  $/$  &        &     \\
  	Video Grid   &  $.384$   & $.106$   &   $/$  &     \\
  	Video Avatar & $***$  &$.900$ &  $*$   & $/$ \\
  	\bottomrule
  \end{tabu}
\end{table}

\emph{Preference.}
The {Friedman Test} shows a significant effect of the four conditions
({$Chi$-$square$  $statistic = 11.89$, $p = .007$}). We further conducted {the Post-hoc Nemenyi Test} for pairwise comparisons and summarized the results in Tab. \ref{tab:Preference}. 
{Compared with \textbf{Video} ($M = 4.2$, $SD = 1.2$), only \textbf{Video Avatar} ($M = 5.5$, $SD = 1.0$) is significantly more preferred. Both \textbf{Avatar} ($M = 4.5$, $SD = 1.5$) and \textbf{Video Grid} ($M = 5.3$, $SD = 1.0$) are not significantly different from it. There are also no significant differences in preference between the rest pairs of conditions.}

\begin{table}[tb]
  \caption{{Post-hoc Nemenyi Test results for Preference in the Evaluation. ($*$ $p < .05$, $**$ $p < .01$, $***$ $p < .001$)}}
  \label{tab:Preference}
  \scriptsize
  \centering
  \begin{tabu}{r*{7}{c}*{2}{r}}
  	\toprule
  	\textbf{Preference} &  Video & Avatar & Video Grid & Video Avatar\\
  	\midrule
  	Video       &  $/$   &       &        &     \\
  	Avatar      & $.900$ &  $/$  &        &     \\
  	Video Grid   & $.122$& $.256$&   $/$  &     \\
  	Video Avatar &  $*$  &$.122$ & $.900$ & $/$ \\
  	\bottomrule
  \end{tabu}
\end{table}

\subsection{Discussion}
\textbf{H1 is supported.} \textbf{Avatar} is perceived to have a significantly lower fidelity than {the} other conditions. Users feel they are less real because ``I cannot see the remote user's facial expression, which is an important clue to tell if he is making up words.'' and ``I cannot tell when the remote user describes his word using gestures and body language''. {This is in line with related research in other scenarios \cite{pan2016comparison,yu2021avatars} and could happen to all widely used avatar-based solutions. However, if the 3D modeling and animation of avatars could be photorealistic enough to cross the uncanny valley in easy-to-set-up configurations in the future, re-evaluation would be necessary. Apple demonstrated its solution on the Vision Pro HMD with life-like scanned upper-body-only avatars\footnote{\url{https://www.apple.com/apple-vision-pro}}. But the real effect is unclear since it is not publicly available yet. Several reviews reflect that, unlike high-end solutions such as Project Starline \cite{lawrence2021project} from Google, it is still uncanny.}

\textbf{H2 is {mainly} supported.} Due to the life-size user portrait ``standing'' on the floor in \textbf{Video Avatar} condition, users perceive remote peers significantly more present than both \textbf{Video} and \textbf{Video Grid} conditions. With the good 3D effect of \textbf{Avatar}, {it feels more present in the physical space than \textbf{Video}. However, the disadvantage in fidelity harms the sense of co-presence to be not significantly better than \textbf{Video Grid}. In addition, the direct transplant on the AR HMD (i.e., \textbf{Video Grid}) cannot bring an improved spatial experience for video conferencing, with no significant difference with \textbf{Video}.}

\textbf{H3 is partially supported.} Compared to \textbf{Video}, \textbf{Video Avatar} is significantly more preferred, but \textbf{Avatar} {and \textbf{Video Grid}} are not. We think it is due to the avatar's drawback of low fidelity {and the similar experience between the direct transplant and conventional video conferencing}. It indicates that fidelity might be more important than co-presence in the conversational scenario. It can also be seen from the insignificant difference in preference between \textbf{Video Grid} and \textbf{Video Avatar}, with \textbf{Video Grid} having a significantly lower co-presence. \textbf{Video Avatar} is the most preferred. Though not significant, its mean score is higher than \textbf{Video Grid}.

In conclusion, the results verify that the video avatar representation can balance fidelity and co-presence, being the better choice for HMD-based small-group AR teleconferencing compared with the existing avatar-mediated approach and the direct AR transplant of video conferencing. Moreover, it can truly upgrade the experience of current video conferencing, being a feasible solution for AR HMDs with its simple setup.

%% file: vrstudy.tex
\section{{Further Study on the Effect of FoV}}
\label{FurtherVRStudy}
In our pilot study, we initially explored the effect of FoV on the video avatar's optimal placement. We conclude {the insignificant results are due to} 
the limited number and range of the FoV conditions. Therefore, to further examine whether such effects exist, we conduct this follow-up study using a VR-simulated AR environment {following existing practice \cite{medeiros2022shielding,evangelista2022auit}}. With OST AR HMDs suffering from limited FoVs, VR simulation can provide a similar experience to Video See-Through (VST) AR HMDs with larger FoV.

For the study condition, other than the \textbf{30-}, \textbf{40-}, and \textbf{50-FoV} conditions {(for comparison with the pilot study to see if there is a difference in the optimal placement between using OST AR HMD and the VR-simulated AR)}, we add \textbf{70-}, \textbf{90-}, and \textbf{110-FoV} conditions {(the range of FoVs of most existing AR and VR HMDs)} in this study. Moreover, we involve the aspect ratio factor in this study. We consider the aspect ratios of the \textbf{16:9} and \textbf{3:2}. Most of the experimental setup is the same as the pilot study, except for using a VR HMD. To include large FoV conditions, we use HTC Vive Pro to present the experience, with an approximate 110$^\circ$ diagonal FoV. We take a 360$^\circ$ picture of the study room and make it the spherical background in the VR environment. We also manipulate the lighting to be more similar to the real environment. We recruited the same group of participants in the pilot study to ensure the design was within-subject. The study scenarios, metrics, and procedures are all the same as in the pilot study. 

\begin{figure}[t]
  \centering
  \includegraphics[width=0.8\columnwidth]{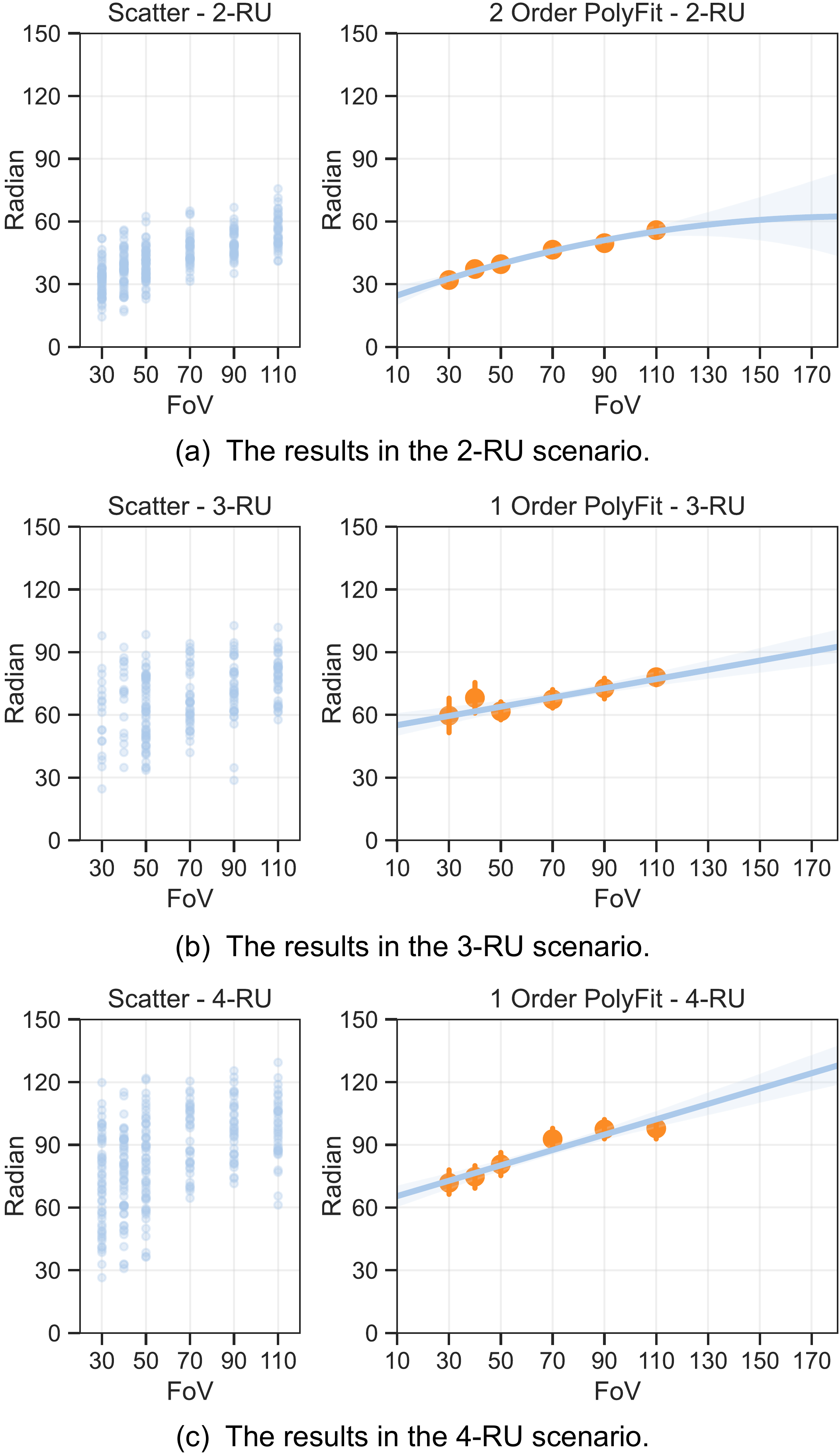}
  \caption{The scatter plots {showing} 
  the original placement data {distribution} and the corresponding polynomial fitting results for $Radian$ in 2-, 3-, and 4-RU scenarios.}
  \label{fig:FittingRadian}
\end{figure}

\subsection{Results}
\emph{The comparison between data collected from AR and VR environments.}
To examine whether the simulated ``AR teleconference scenario'' can actually represent the true AR experience, we {first} compare the $Radian$ and $Radius$ data with the same $FoV$s (30$^\circ$, 40$^\circ$, and 50$^\circ$) and $aspect$ $ratio$ ($3:2$) from our experiments in VR simulation and AR environment to see if there are significant differences. For the result analysis, we examined the normality of the data using Shapiro-Wilk Normality Tests and conducted pairwise T-Tests on normally distributed samples and non-parametric {Wilcoxon Signed-Rank Tests} 
on non-normally distributed samples for each scenario, adding up to 21 pairs of data to test in total. {We also ran TOST on these pairs to initially test the equivalence between these two groups of data. We set the TOST bound to 20 for $Radian$ and 0.5 for $Radius$.} The comparisons are visualized in Fig. 2 in the supplemental file, 
and the significance test results {and the initial equivalence test results} are shown in Tab. 2 in the supplemental file. Other than two exceptions, which will be illustrated later, we found the rest 19 pairs of data not significantly different from each other within pairs. The two exceptions both occur on $Radian$ in the 3-RU Scenario. The $Radian$ in $30$-$FoV$ condition in AR ($M = 59.6$, $SD = 18.3$) is significantly larger than that in VR ($M = 46.7$, $SD = 13.3$), $t$-$statistic = 2.36$, $p = .024$, {$TOST p = .106$}. The $Radian$ in $40$-$FoV$ condition in AR ($M = 68.1$, $SD = 16.8$) is also significantly larger than that in VR ($M = 49.2$, $SD = 14.8$), $t$-$statistic = 3.50$, $p = .001$, {$TOST p = .418$}. {We discarded the VR data from these two exceptions and combined other data that is not significantly different in VR and AR studies for the subsequent correlation and regression analyses.} We discuss the results and the two exceptions in Sec. \ref{VRDiscussion}. For other statistical results showing {the significant equivalence and} no significant difference in data between AR and VR environments, please refer to Tab. 2 in the supplemental file.


\emph{The comparison of the aspect ratio.}
{We also conducted pairwise T-Tests and non-parametric Wilcoxon Signed-Rank Tests on normally and non-normally distributed samples, respectively.} The comparison of the data and the statistical results are shown in Fig. 1 and Tab. 3 in the supplemental file. We can see that the 16:9 and 3:2 aspect ratios do not have a significant effect on the placement of remote users in video-avatar-based AR teleconference applications {in most conditions}. 

\emph{Correlation test.} To examine whether there are correlations between the remote user placement (formulated by the $Radian$ and $Radius$ values) and the $FoV$, we first ran normality tests to determine the normality of the data and then conducted Pearson Correlation Test and Spearman Correlation Test on the normally distributed samples and the non-normally distributed samples, respectively. We summarized the corresponding statistical results in Tab. \ref{tab:CorrelationVR}. From the correlation test results, we can see that both $Radian$ and $Radius$ are significantly correlated with $FoV$. The correlation between $Radian$ and $FoV$ is positive while the correlation between $Radius$ and $FoV$ is negative. 

\begin{table}[tb]
  \caption{{Correlation Test results between the $Radian$ and $Radius$ values and the $FoV$ in the VR Study. ($*$ $p < .05$, $**$ $p < .01$, $***$ $p < .001$)}}
  \label{tab:CorrelationVR}
  \scriptsize
  \centering
  \begin{tabu}{r*{7}{c}*{2}{r}}
  	\toprule
        &           & 1-RU   & 2-RU   & 3-RU  & 4-RU  \\ 
        \midrule
        \multirow{2}{*}{Radian} & Statistic & $/$  & \begin{tabular}[c]{@{}c@{}}Pearson\\ .69\end{tabular}   & \begin{tabular}[c]{@{}c@{}}Pearson\\ .36\end{tabular}   & \begin{tabular}[c]{@{}c@{}}Spearman\\ .46\end{tabular}  \\
        & $P$-Value   & $/$  & $***$    & $***$     & $***$   \\
        \multirow{2}{*}{Radius} & Statistic & \begin{tabular}[c]{@{}c@{}}Spearman\\ -.36\end{tabular} & \begin{tabular}[c]{@{}c@{}}Spearman\\ -.32\end{tabular} & \begin{tabular}[c]{@{}c@{}}Spearman\\ -.29\end{tabular} & \begin{tabular}[c]{@{}c@{}}Spearman\\ -.24\end{tabular} \\
        & $P$-Value   & $***$   & $***$     & $***$       & $***$   \\
  	\bottomrule
  \end{tabu}
\end{table}

\begin{figure}[t]
  \centering
  \includegraphics[width=0.8\columnwidth]{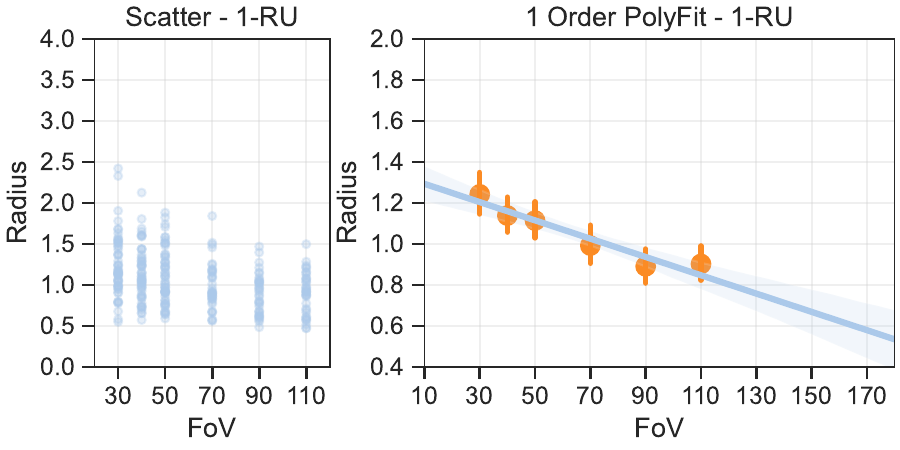}
  \caption{%
  	The scatter plots showing the placement data distribution and the corresponding polynomial fitting results for $Radius$ in the 1-RU scenario.
  }
  \label{fig:FittingRadius-1RU}
\end{figure}

\begin{figure}[t]
  \centering
  \includegraphics[width=0.8\columnwidth]{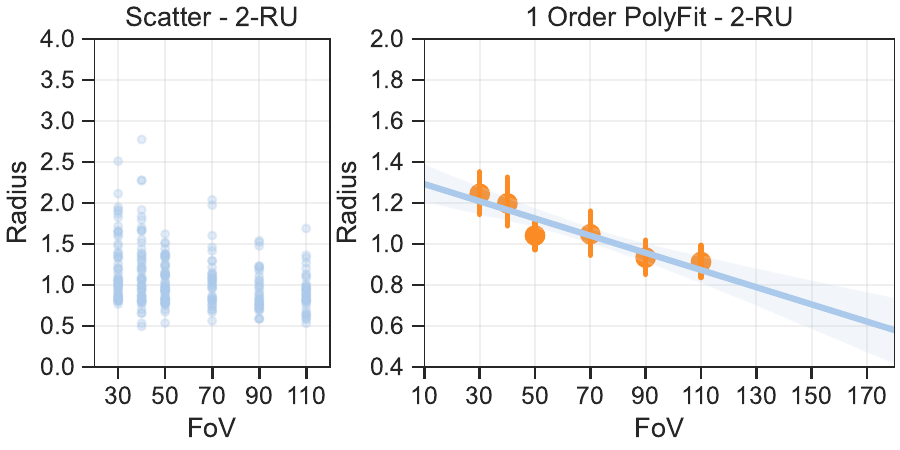}
  \caption{%
  	The scatter plots showing the placement data distribution and the corresponding polynomial fitting results for $Radius$ in the 2-RU scenario.
  }
  \label{fig:FittingRadius-2RU}
\end{figure}

\begin{figure}[t]
  \centering
  \includegraphics[width=0.8\columnwidth]{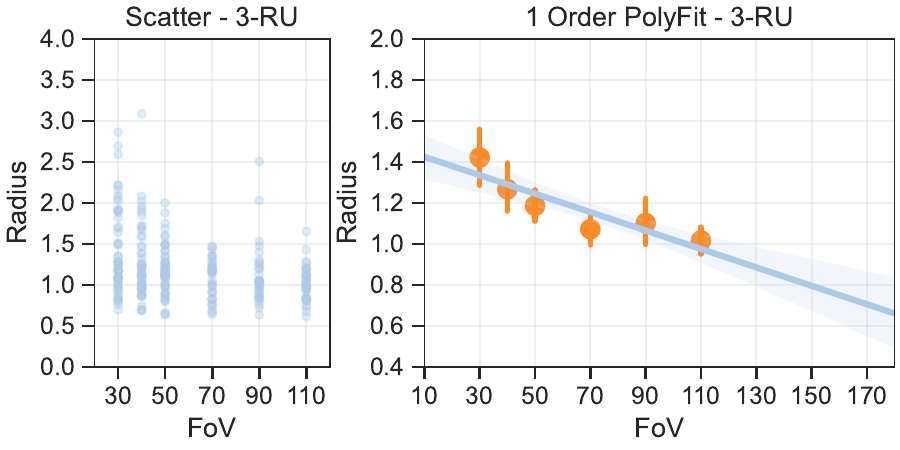}
  \caption{%
  	The scatter plots showing the placement data distribution and the corresponding polynomial fitting results for $Radius$ in the 3-RU scenario.
  }
  \label{fig:FittingRadius-3RU}
\end{figure}

\begin{figure}[t]
  \centering
  \includegraphics[width=0.8\columnwidth]{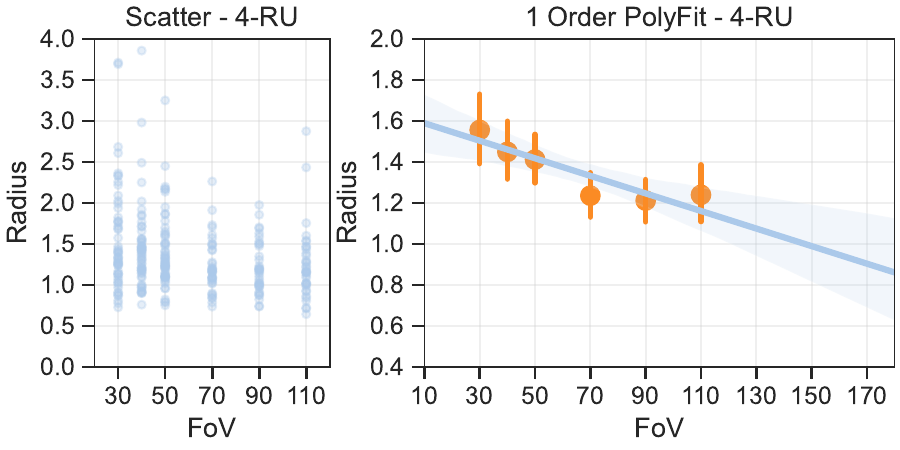}
  \caption{%
  	The scatter plots showing the placement data distribution and the corresponding polynomial fitting results for $Radius$ in the 4-RU scenario.
  }
  \label{fig:FittingRadius-4RU}
\end{figure}

\emph{Model fitting.} With the significant correlation results, we can now {quantitatively} explore the mapping from $FoV$ to $Radian$ and $Radius$. We regressed the data in each scenario using Polynomial Fitting with $FoV$ as the independent variable. The interview {conducted} after the participants finished all tasks indicates that their adjustment of the $Radian$ and $Radius$ was consistent and monotonic. We thus choose the highest order for each fitting function that results in a monotonic curve when $FoV$ = [0, 180]. We present the distribution of the data and the fitting results for $Radian$ and $Radius$ in Figs. \ref{fig:FittingRadian},\ref{fig:FittingRadius-1RU},\ref{fig:FittingRadius-2RU},\ref{fig:FittingRadius-3RU},\ref{fig:FittingRadius-4RU}, respectively. We choose $FoV$ as the only independent variable and set the $number$ $of$ $remote$ $users$ as a scenario parameter instead of another independent variable because our application targets the small-group conversation use case and shapes the group as a circle. With the increase in participant numbers, the group formation could change to layouts with multiple rows or layers and may vary regarding the functional area and scene semantics. Therefore, a model regressed from the independent variable containing only small participant numbers would be less descriptive in scenarios involving large numbers of participants, in which case separately building new models is necessary. From the regression result, we can see that in order to meet the monotone constraints described earlier, except the $Radian$ model in the 2-RU scenario was regressed as a quadratic function (as shown in Eq. \ref{eq:Radian2RU}, {$F(2, 267) = 124.5$, $p < .001$}), 
the rest sets of data were regressed linearly. They are formulated as $\lambda_0 FoV + \lambda_1$ with $\lambda_0$ and $\lambda_1$ denoting the first and second values in the coefficients respectively {and $FoV$ denoting the $FoV$ value}. We list the values of the coefficients {and $F$-Test results} 
in Tab. \ref{tab:OtherModels}. 
{Note that the difference in the $df$ of the $F statistic$ in the 3-RU scenario for $Radian$ is caused by the removal of exceptional data as illustrated earlier.}

\begin{equation}
  \label{eq:Radian2RU}
  Radian_{2RU} = -0.0012{FoV}^2 + 0.45{FoV} + 20.20.
\end{equation}

\begin{table}[tb]
  \caption{{Coefficients and $F$-Test results 
  of linear regressed models in the VR Study. ($*$ $p < .05$, $**$ $p < .01$, $***$ $p < .001$)}}
  \label{tab:OtherModels}
  \scriptsize
  \centering
  \begin{tabu}{r*{7}{c}*{2}{r}}
  	\toprule
        &           & 1-RU   & 2-RU   & 3-RU  & 4-RU  \\ 
        \midrule
        \multirow{3}{*}{Radian} & Coefficients & $/$  & $/$   & \begin{tabular}[c]{@{}c@{}}0.22,\\ 52.87\end{tabular}   & \begin{tabular}[c]{@{}c@{}}0.37,\\ 61.79\end{tabular}  \\
        & $F$ & $/$ & $/$ & \begin{tabular}[c]{@{}c@{}}$F(1, 196)$\\$= 29.38$\end{tabular}  & \begin{tabular}[c]{@{}c@{}}$F(1, 268)$\\$= 73.40$\end{tabular} \\
        & $P$-Value  & $/$ & $/$ & $***$ & $***$ \\
        \\
        \multirow{3}{*}{Radius} & Coefficients & \begin{tabular}[c]{@{}c@{}}-0.0045,\\ 1.34\end{tabular} & \begin{tabular}[c]{@{}c@{}}-0.0042,\\ 1.33\end{tabular} & \begin{tabular}[c]{@{}c@{}}-0.0045,\\ 1.47\end{tabular} & \begin{tabular}[c]{@{}c@{}}-0.0043,\\ 1.63\end{tabular} \\
        & $F(1, 268)$ & 39.84 & 29.62 & 27.94 & 15.63 \\
        & $P$-Value  & $***$ & $***$ & $***$ & $***$ \\
  	\bottomrule
  \end{tabu}
\end{table}

\subsection{Discussion\label{VRDiscussion}}
{\textbf{Choice of aspect ratio.} The two conditions in our study are not significantly different from each other.} However, {some} participants reflect that ``When the FoV decreases vertically, I would let the remote users stand a little further away from me to see more parts of each user's body than rotate my head vertically to observe''. ``Vertically rotating my head to see more body parts of a person feels impolite. It is like I am checking out him/her from head to toe while he/she is standing at a rather close distance during the conversation''. ``Wearing an HMD makes it tiring to rotate my head vertically''. These reflections, along with the relatively smaller mean $Radius$ values for the 3:2 aspect ratio in most of the conditions, {imply} that for video-avatar-based AR teleconference applications, when the FoV is limited by hardware, a lower horizontal-vertical aspect ratio might be {a} 
better choice. We can further explore lower aspect ratios such as 4:3, 4:5, 2:3, and 9:16 to determine the optimal aspect ratio for video-avatar-based AR teleconference applications. 

\textbf{Validity of VR simulation.} {It is common in previous research \cite{ren2016evaluating,medeiros2022shielding,evangelista2022auit} 
to adopt VR simulations to explore AR problems. We initially verify its validity in our small group teleconferencing scenarios by finding no} significant difference {and significant equivalence} between most of the data from our experiment in VR simulation and AR environment. {It makes sense} to explore our video-avatar-based AR teleconferencing application in bigger-FoV conditions in the VR-simulated AR environment when AR HMDs currently available only provide rather limited FoVs. {Further specific tests to compare studying in VR simulation and directly on AR devices could be useful regarding the empirical bound values with larger data collections.}

\textbf{The results are applicable to both VST and OST AR HMDs.} 
Our VR-simulated AR environment provides an experience similar to the VST AR HMD in terms of the HMD's size, weight, wearing {comfort}, 
and imaging quality. Together with the experiment in the AR environment using an OST HMD (i,e, Microsoft HoloLens 2), our study covers these two major types of devices from the perspective of overall user experience. Therefore, {we initially reason that our} results are applicable to both situations. {Specific further quantitative studies would be useful to study the difference between these two types of AR HMDs or even among various products.}

\textbf{The two exceptional cases.} The $Radian$s in \textbf{30-} and \textbf{40-FoV} conditions in the 3-RU scenario are significantly larger in AR than in VR. This can be explained using the reflection from some participants: ``I am more willing to rotate my head when wearing the AR HMD because it is much lighter,'' and ``Rotating the head in AR feels easier than in VR because there is no {motion} sickness issue''. Note that the ``AR HMD'' and ``AR'' in the quotations both refer to the OST AR HMD used in our pilot study. We think the reasons why the exceptions occur only in the \textbf{30-} and \textbf{40-FoV} conditions in the 3-RU scenario are that 1) very small FoVs lead to significantly more head rotations during the study, which makes the difference between AR and VR significant as well. 2) Participants want to adjust the video avatars' placement while observing every one of them in 1-, 2-, and 3-RU scenarios, while {they} tend to only refer to two or three video avatars when doing the task in the 4-RU scenario. This makes 3-RU the scenario requiring the most head rotations since observing the whole RU group in 1- and 2-RU scenarios and the partial RU group in the 4-RU scenario is easier with targets distributed in relatively small angle ranges. We drop the VR data in these two exceptional scenarios in the model regressing process.

\textbf{The effectiveness of the regressed model.} When the HMD's FoV can cover the full FoV of human eyes with $FoV$ = 180, the values predicted by our models are 
in line with those in real scenarios. $Radian$ approaches the inside angles of regular polygons with vertices being the participants in the group conversation (60 for a triangle, 90 for a rectangle, and 120 for a pentagon, corresponding to the 2-, 3-, and 4-RU scenarios). $Radius$ in each of the four scenarios with different numbers of users also reaches the value that can form a layout that preserves the common distance for friends or strangers (around 1m \cite{hall1968proxemics,hecht2019shape}) between each participant. This similarity between the predicted layout and the common real-world layout also proves the effectiveness of our regressed models and the overall high level of realism and co-presence provided by the video-based avatars.

%% file: conclusion.tex
\section{Conclusion, Limitations, and Future Work}
In conclusion, we have explored the use of {{life-size} 2D} video avatars to represent remote users in small-group AR teleconferencing on the HMD platform. We have conducted a pilot study {first to} 
find the optimal placement of video avatars for teleconferencing on {HoloLens 2}, one of the most commonly used AR HMDs. We have also found initial implications of the effect of FoV on the video avatars' optimal placement in the pilot study. To demonstrate the use of video avatars in a real-time AR teleconferencing system, we have developed a proof-of-concept prototype {and conducted an evaluation using the prototype and the optimal placement obtained from the pilot study. We found that using video avatars achieves a balance between fidelity and co-presence in HMD-based AR teleconferencing, and outperforms conventional videoconferencing}. 
To make the video avatars' optimal placement results more {generalizable} 
and applicable to future AR HMDs with bigger FoVs, we have conducted a further study in a VR-simulated AR environment and fitted placement models as {quantitative} references for the teleconferencing applications on both current-available and future AR HMDs. {Our findings through the series of user studies along with the proof-of-concept prototype could potentially benefit the development of widely accessible and immersive HMD-based AR telepresence applications.} Next, we {discuss} 
the limitations of our work and the corresponding potential future work.

{\textbf{Diversity}. Our current studies involve limited participant diversity. We are interested in conducting larger-scale studies with a more diverse population in gender, age, race, and appearance to make the results more generalizable.}


\textbf{Constrained point of view}. 
This is the intrinsic limitation of the 2D video representation. We currently {rotate the billboard vertically to make it}
always orient to the user. As a result, if the local user moves around, he/she still can only see the remote users from the frontal view. Moreover, though we ensure the video avatars are life-size, users of different heights still have different points of view vertically. To tackle this problem, we will 
explore incorporating neural rendering techniques \cite{mildenhall2021nerf,pumarola2021dnerf,gafni2021dynamic} to synthesize and show the correct view according to the user's head position in real-time.

\textbf{Semantic interaction with physical environments}. We currently only focus on the standing pose and make sure the video avatars' feet are on the floor in our prototype system. In the future, we will explore more poses (e.g., sitting and leaning) and align video avatars with objects that are semantically affordable for the pose. Additionally, when a remote user moves around in the remote space, we need to adapt the position and content of his/her video avatar regarding the semantics of the physical space as existing approaches do with virtual avatars \cite{kim2016retargeting,tahara2020retargetable,yoon2020placement,wang2022predict}. We can also adjust the lighting and shadowing effects in the video to adapt to the lighting condition in the local environment \cite{song2022real}. {In addition, the influence of presenting lighting and shadowing effects that match the physical environment for video avatars could be further studied.}

{\textbf{Incorporating more features}. We take the first step to investigate the use of video avatars in our current studies utilizing a rather simple proof-of-concept prototype. Adding more features such as visualizing shared gaze, user attention, and gestures \cite{piumsomboon2018mini} could further resemble a stronger system. We can further adapt such existing visualization techniques to our video-avatar-based prototype.}

\textbf{More integrated setup}. We use a webcam put statically about 2 meters away at 
approximately the same height as the remote user's eyes to capture the whole body portrait. With the miniaturization of the commodity 360$^\circ$ cameras and image distortion correction techniques \cite{elgharib2020egocentric}, we can explore more integrated configurations such as attaching a 360$^\circ$ camera to the HMD to capture the user.

\textbf{Large scale scenarios}. In this work, we focus on the small-group teleconferencing scenario involving up to 5 participants (counting all the remote and local users). We can further explore scenarios with tens and even hundreds of people to apply the application for use cases such as hybrid classrooms and virtual events.